\documentclass{article}
\usepackage[table]{xcolor}
\usepackage{longtable}
\usepackage[letterpaper, total={6.0in, 9in}]{geometry}
\usepackage{graphicx}
\usepackage{caption}
\usepackage{amsmath,amssymb}
\usepackage{authblk}

\title{Triply-resonant coupled-cavity electro-optic modulators for on-chip RF photonic systems}

\author[1,*]{Hayk Gevorgyan}
\author[1]{Anatol Khilo}
\author[2]{Yossef Ehrlichman}
\author[1,2]{Milo\v s Popovi\'c}

\affil[1]{Department of Electrical and Computer Engineering, Boston University, Boston, MA 02215, USA}
\affil[2]{Department of Electrical, Computer, and Energy Engineering, University of Colorado, Boulder, CO 80309, USA}
\affil[*]{hayk@bu.edu} 

\date{}

\begin{document}

\maketitle

\begin{abstract}
We propose a triply-resonant electro-optic modulator architecture which maximizes modulation efficiency using simultaneous resonant enhancement of the RF drive signal, the optical pump, and the generated optical sideband. Optical enhancement of the optical pump and the sideband is achieved using resonant supermodes of two coupled optical resonators, and the RF enhancement is achieved with LC circuits formed by the capacitances of the optical resonators and inductors which can be implemented using CMOS technology. In the proposed configuration, the photon lifetime determines the bandwidth of the RF response but not its center frequency, which is set by the coupling strength between the two resonators and is not subject to the photon lifetime constraint inherent to conventional single-mode resonant modulators. This enables efficient operation at high RF carrier frequencies without a reduction in efficiency commonly associated with the photon lifetime limit. Two optical configurations of the modulator are proposed: a ``basic'' configuration with equal Q-factors in both supermodes, most suitable for narrowband RF signals, and a ``generalized'' configuration with independently tailored Q-factors of the two supermodes, which makes it possible to broaden the RF bandwidth without sacrificing the resonant enhancement of the optical pump and paying a penalty in modulation efficiency associated with doing so. Additionally, a significant gain in modulation efficiency is expected from RF signal enhancement by LC resonant matching. This results in a modulator which is compact, efficient, and capable of modulation at high RF carrier frequencies. The proposed modulator architecture optimally engineers the interaction of the device with each of the three signals and between them, can be applied to any modulator cavity design or modulation mechanism and promises to enable complex RF-photonic systems on chip.
\end{abstract}

\section{Introduction} \label{sec:Intro}
Microwave photonic (MWP) systems rely on sensitive electro-optic (EO) modulators for various applications such as radio-over-fiber, optical beam forming, photonic signal processing, photonic analog-to-digital conversion, satellite-based mm-wave sensing, etc. \cite{Capmany2007,Marpaung2013,Minasian2006,Pett2018}. The efficiency with which the RF signal is converted to the optical domain by an EO modulator is an essential parameter directly affecting the gain of a MWP link. 

Mach-Zehnder (MZ) modulators, both discrete and integrated, have been used in MWP systems over the past decades as the workhorse devices for RF-to-optical conversion. However, these devices are large and power-hungry. Integrated photonics technology offers new opportunities for sensitive microresonator-based EO modulators, such as microring and photonic crystal cavity modulators. Owing to their resonant nature, these devices are compact and efficient. However, they suffer from the speed-sensitivity tradeoff, imposed by the cavity photon lifetime \cite{Williamson2001,poon2008,Ehrlichman2018}. This tradeoff has been addressed by using modulation-induced coupling between adjacent free spectral range (FSR) modes of millimeter-scale whispering-gallery-mode disk and ring resonator modulators, which have their speed set by the FSR of the resonator \cite{levi2001,maleki2003,lipson2014}, as discussed in more detail below. These devices are inherently large and require implementation of RF traveling wave electrodes. In this work, we present a detailed account of a novel triply resonant modulator consisting of two coupled microresonator-based modulator cavities and lumped RF resonators. The proposed device is compact and energy efficient, similar to conventional microresonator-based modulators. Moreover, it does not suffer from speed-sensitivity tradeoff, since it uses multiple optical resonances, similar to FSR-coupled modulators. Additionally, the RF resonance boosts the voltage which further increases the efficiency, as described below.

\begin{figure}[!t]
\centering
\mbox{\includegraphics{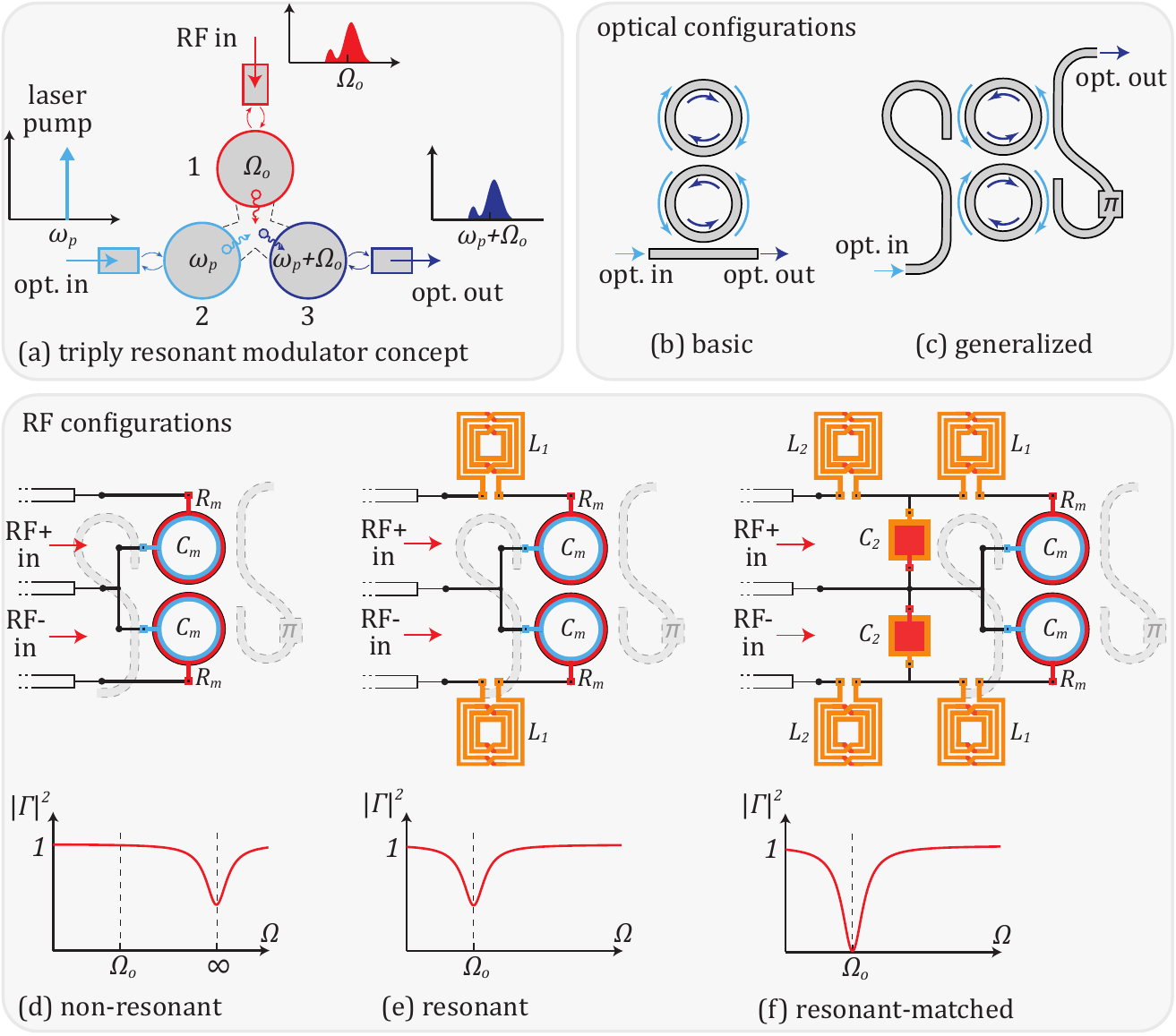}}
\captionsetup{width=.9\textwidth,font=small}
\caption{(a) Conceptual representation of the triply-resonant modulator, consisting of one RF and two optical mutually coupled resonances at frequencies of the three interacting waves: the input RF drive, the input laser pump, and the output optical sideband. (b,c) Physical realization of the two optical resonances by supermodes of (b) the basic coupled-cavity design, with a conventional bus waveguide, and (c) the generalized coupled-cavity design, with novel interferometrically coupled input/output waveguides. (d) Non-resonant RF design, where the transmission line directly feeds the active optical cavities. Due to impedance mismatch between them, RF power is almost completely reflected from the load. (e) Resonant design, where the RF resonance from (a) is realized by LC circuits, consisting of integrated inductors $L_1$ and capacitances $C_m$ of the active cavities. It greatly reduces RF power reflection and enhances the voltage on the active cavities. (f) Resonant-matched design, where critical coupling between the transmission line and the RF resonator is achieved using impedance matching circuits which consist of inductors $L_2$ and capacitors $C_2$. The plots at the bottom of (d)-(f) sketch the frequency dependence of the power reflection $\lvert\Gamma\rvert^2$ of the RF signal back into the transmission line.}
\label{fig:GenModOptRFConf}
\end{figure}

\subsection{The proposed modulator} 
This work proposes a novel electro-optic modulator for efficient conversion of RF signals into optical domain. The RF signals are assumed to be bandlimited with RF carrier frequency $\Omega_o$. An electro-optic modulator can be viewed as a device where the interaction of two (input) waves, the laser pump wave and the RF drive wave, creates a third (output) wave -- an optical sideband shifted in frequency by $\Omega_o$ with respect to the laser pump. The goal is to maximize the modulation efficiency, defined here as the fraction of the input laser pump power which gets converted into the optical sideband (see Sec.~\ref{sec:formulation} for formal definition).

This work proposes an integrated triply-resonant RF electro-optic modulator, where each of the two optical waves and the RF wave are resonantly enhanced to maximize the modulation efficiency. Our work is a continuation of that in \cite{wade2014}, and some aspects of this modulator have been recently described in \cite{Ehrlichman2016, Ehrlichman2017, loncar2017, Gevorgyan2018}. A conceptual representation of the triply-resonant device is shown in Fig.~\ref{fig:GenModOptRFConf}(a). The three rectangles represent the input/output ports of the device, and the three circles represent the three mutually coupled resonances. The RF resonance and one of the optical resonances are excited by the input RF and the laser pump waves, while the other optical resonance is excited by the optical sideband that is generated due to nonlinear interaction between the RF drive and the optical pump within the device. During this interaction, an RF photon combines with a pump photon to produce an optical sideband photon, translating each RF spectral component to the optical domain. The key idea of the proposed device is that the conversion efficiency is maximized when all three interacting waves are at resonance, and when their lifetimes/escape efficiencies are properly tailored.

In the physical world, the abstract triply-resonant modulator of Fig.~\ref{fig:GenModOptRFConf}(a) can be realized with two coupled optical cavities connected to an RF resonator. The resonance frequencies of the two optical cavities can be tuned by applying a voltage to their built-in capacitive phase shifters, which creates coupling between the optical and the RF waves. The phase shifters can be based e.g. on the carrier plasma or linear electro-optic Pockels effects. In the text below, we will refer to these tunable optical cavities as ``\textit{active cavities}''.

The \textit{two optical resonances} of the abstract device of Fig.~\ref{fig:GenModOptRFConf}(a) correspond to the two resonant supermodes of the coupled optical cavities, as described below. The two optical cavities are identical and have the same resonance frequencies when uncoupled. In the proposed device, the two cavities are evanescently coupled as illustrated in Figs.~\ref{fig:GenModOptRFConf}(b,c). The coupling produces two new orthogonal states -- the symmetric and antisymmetric supermodes -- with resonance frequencies split due to the coupling. The coupling strength is selected to ensure that the frequencies of the symmetric and antisymmetric supermodes are separated by the RF carrier frequency $\Omega_o$. The two supermodes of the coupled cavities correspond to the two optical resonances of the abstract device in Fig~\ref{fig:GenModOptRFConf}(a). The input pump laser is matched in frequency to the symmetric supermode, while the frequency of generated optical sideband equals the frequency of the antisymmetric supermode (or vice versa). This work considers single sideband generation, but can be easily extended to dual sideband generation in a three-optical-resonance system following \cite{wade2014}.

This work considers two configurations of the coupled optical cavities: the \textit{basic coupled-cavity design} shown in Fig.~\ref{fig:GenModOptRFConf}(b), and the \textit{generalized coupled-cavity design} shown in Fig.~\ref{fig:GenModOptRFConf}(c). In the basic configuration, both supermodes have the same quality factor (escape efficiency), while the generalized configuration enables independent control of the Q-factors of the optical resonances, leading to higher modulation efficiency for broadband RF signals, as described in Sec.~\ref{sec:OptResAnalysis}.

The \textit{RF resonance} in the abstract device of Fig.~\ref{fig:GenModOptRFConf}(a) is introduced by two LC resonators formed by integrated inductors and the capacitance of the electro-optic region of the active optical cavities. As shown in Fig.~\ref{fig:GenModOptRFConf}(e), the two LC resonators are identical with equal resonance frequencies and are connected to RF transmission lines. The transmission line delivers a differential RF signal, driving the two active cavities in push-pull mode. Unlike the optical resonators which are coupled, the two RF resonators are uncoupled, and their resonances are degenerate and behave as a single resonance, which is the RF resonance depicted in Fig.~\ref{fig:GenModOptRFConf}(a). The frequency of this resonance is matched to the RF carrier frequency $\Omega_o$. If the RF drive is directly applied to the capacitances of the active cavities [Fig.~\ref{fig:GenModOptRFConf}(d)], a significant fraction of the RF power is reflected due to impedance mismatch. To a large extent, the reflection can be mitigated by connecting integrated inductors in series with the capacitances of the active cavities as shown in Fig.~\ref{fig:GenModOptRFConf}(e), forming LC circuits with resonance frequencies equal to the RF carrier. The LC circuit removes the reactive load from the transmission line and boosts the voltage on the active cavities. Perfect impedance matching between the transmission line and the modulator is still not guaranteed, because the transmission line impedance (usually 50 Ohm) is not necessarily equal to the parasitic resistance of the active cavities. By introducing a lumped-element impedance matching circuit \cite{Lee2004}, as shown in Fig.~\ref{fig:GenModOptRFConf}(f), critical coupling between the transmission line and the optoelectronic LC resonator can be achieved. This maximizes the field in the capacitive EO region for a given RF drive power. The remainder of work analyzes all of the RF configurations shown in Figs.~\ref{fig:GenModOptRFConf}(d-f).

\begin{figure}[!t]
\centering
\mbox{\includegraphics{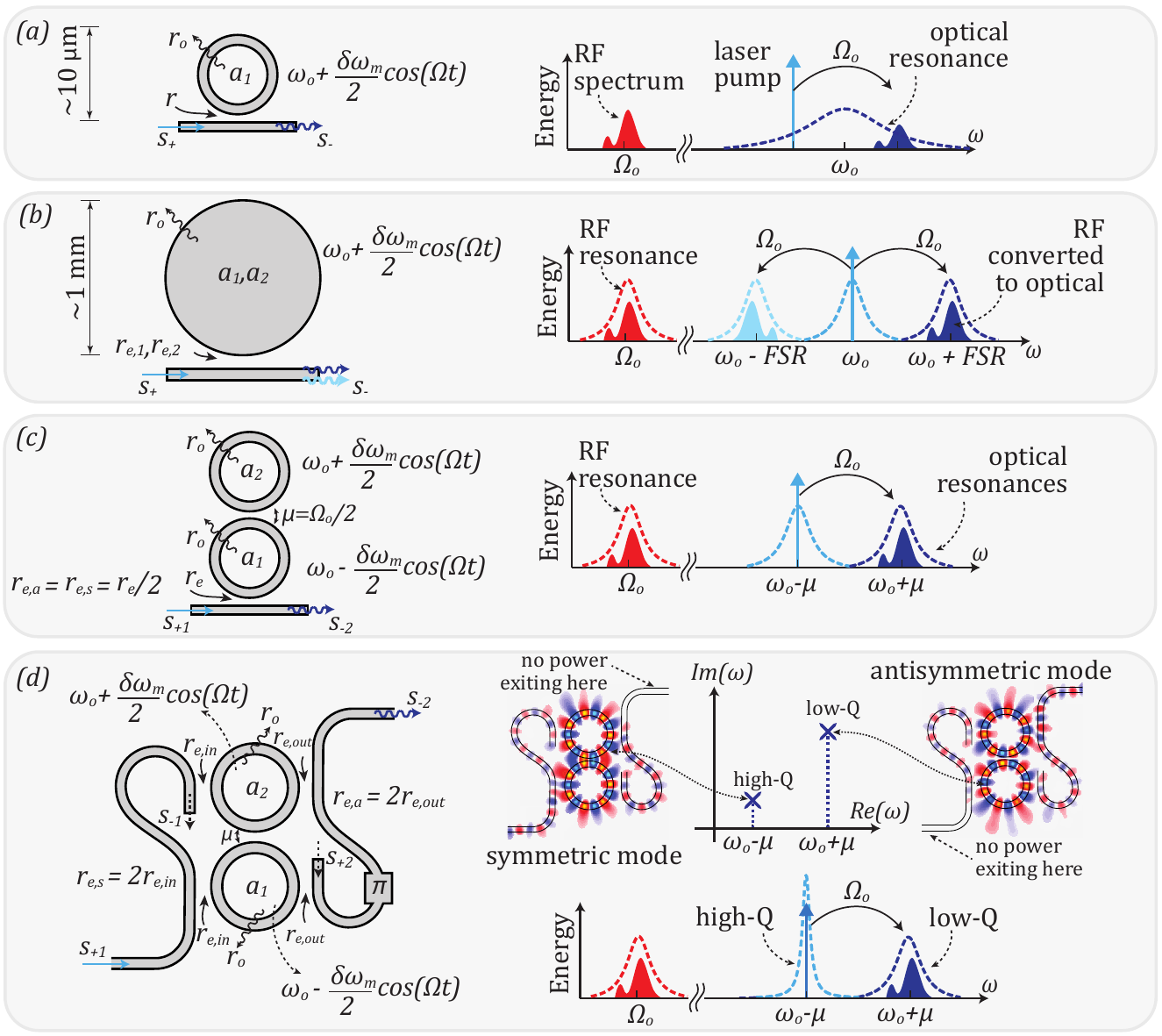}}
\captionsetup{width=.9\textwidth,font=small}
\caption{Resonant modulators for RF-to-optical conversion (a,b) previously demonstrated and (c,d) studied in this work. (a) Regular microring modulators are compact and efficient, but suffer from the speed-efficiency tradeoff. (b) Millimeter-scale disk or ring resonator modulators, which overcome the speed-efficiency tradeoff through modulation-induced coupling between multiple resonant mode orders at adjacent FSRs, have large footprint and require implementation of RF traveling-wave electrodes. (c,d) Optical coupled-cavity designs of the proposed modulator perform RF-to-optical conversion by transferring optical energy from the symmetric to the antisymmetric supermode resonance, via push-pull modulation of the resonance frequencies of the coupled microresonators. The designs are compact, efficient and do not suffer from the speed/efficiency tradeoff. Additionally, unlike the basic proposed design, where supermodes have equal Q-factors, (d) the generalized design allows independent tailoring of supermode Q-factors, providing higher efficiency with larger modulation bandwidth if the symmetric resonance is kept at critical coupling and the antisymmetric resonance is broadened just enough to accommodate the RF spectrum.}
\label{fig:ModTypes}
\end{figure}

The proposed modulator can in principle be implemented in any material platform that allows integration of the optical and the RF resonators. Especially attractive are high-index-contrast waveguide platforms with efficient phase shifter mechanisms, with several metal layers available for building integrated inductors. With tight confinement of the optical and RF fields in the high-index-contrast active cavities and RF resonant enhancement, such platforms are promising candidates for implementing compact and efficient modulators based on the proposed concept. 

In particular, we are interested in realization of the proposed concept in advanced CMOS processes, such as ``zero-change'' monolithic electronic-photonic 45 nm CMOS \cite{Stojanovic2018}. Integration of silicon photonic devices and circuits in advanced RF CMOS processes (some having standard $f_T$/$f_{max}$ of 305/380 GHz \cite{45rfsoi2017} and $f_T$ as high as 485 GHz \cite{Lee2007}), in close proximity to RF-electronic circuits, bears potential for realization of large-scale, on-chip MWP systems, such as RF-optical beamforming networks for next generation mobile communication systems, satellite-based mm-wave sensors for atmospheric temperature sounding \cite{Pett2018}, etc. Demonstrations of high-speed microring modulators on an advanced digital CMOS process \cite{Stojanovic2018}, which has been recently geared toward millimeter-wave and 5G applications \cite{45rfsoi2017}, and the availability of high-Q inductors in the process design kit pave a direct path to realization of the proposed modulator.

\subsection{Comparison to prior art}
Over the past decades MZ modulators have been extensively used in MWP systems, both in discrete component and photonic integrated circuit form. Broadband MZ modulators with up to 40~GHz bandwidth have been demonstrated in silicon \cite{plant2015,hochberg2014}. Modulation beyond 100~GHz has been achieved with MZ modulators based on lithium niobate, electro-optic polymer, and plasmonic waveguides \cite{miyazawa1998,mcgee2002,leuthold2017}. MZ modulators, employing traveling-wave electrodes, overcome bandwidth limitations due to the RC time constant and provide impedance matching with the input feed line. Nevertheless, the bandwidth of these devices does have limits imposed by imperfect velocity matching between the RF and optical waves which has stronger impact on the bandwidth of longer devices. Therefore, there is a tradeoff between efficient modulation and large bandwidth, unless near-perfect velocity matching is achieved.

Microring modulators (based on a single ring, referred to here as ``regular'' microring modulators), shown in Fig.~\ref{fig:ModTypes}(a), take advantage of optical field enhancement in the cavity through multiple round trip propagation \cite{Xu2005}. When implemented in high-index-contrast materials, these devices feature small size and tight spatial confinement of optical and RF electric fields. The small size of the device eliminates the need for traveling-wave electrodes and permits a large RC-time-limited bandwidth. In silicon photonics, microring modulators have been extensively used for low-energy data modulation \cite{moazeni2017,watts2014} and have enabled energy-efficient photonic interconnects \cite{sun2015}. Modulation frequencies up to 40~GHz have been achieved \cite{watts2014,campenhout2015}. Similar to MZ modulators, regular microring modulators suffer from a tradeoff between the modulation speed and efficiency -- in this case that is inherent to singly-resonant devices. This tradeoff arises due to the cavity photon lifetime: increasing the photon lifetime helps to improve the efficiency but limits the modulation speed, while decreasing the photon lifetime makes the modulator faster, but reduces the efficiency \cite{Williamson2001,poon2008,Ehrlichman2018}. The design of regular microring modulators for RF applications has been studied in \cite{Ehrlichman2018}.

Modulation at RF carrier frequencies extending far beyond resonance linewidth of the modulator has been demonstrated by using different resonant modes of millimeter-scale whispering-gallery mode disk or ring resonators. In these devices, the RF carrier frequency is matched to the FSR of the resonator, and RF modulation couples adjacent longitudinal resonant modes which are spaced in frequency by the FSR, transferring pump light from one resonance to another, as illustrated in Fig.~\ref{fig:ModTypes}(b) \cite{levi2001,maleki2003,lipson2014}. We refer to such modulators as ``FSR-coupled'' modulators. This approach eliminates the tradeoff between the modulation speed and efficiency, and has been developed for lithium niobate (LN) whispering-gallery mode disk modulators which exhibit very high quality factor $(\sim 10^6-10^9)$. These modulators have found application as RF-optical mixers in photonic receivers that provide high sensitivity and large dynamic range \cite{levi2001,maleki2003}. However, due to the large cross-sectional area of the waveguide and the optical mode, RF electrodes need to be spaced farther from the waveguide core, resulting in weak RF electric field in LN and limiting the coupling strength between the optical and microwave fields. Additionally, the requirement to have the resonator FSR equal to the microwave carrier frequency, typically in the range of 1 to 100 GHz, leads to a large size of the FSR-coupled resonators (respective radius of 10~mm to 100~$\mu$m for high-index-contrast materials), and necessitates the use of traveling- or standing-wave RF transmission line electrodes\cite{lipson2014}.

The modulator studied in this work incorporates the best features of the regular microring modulator and the FSR-coupled modulator shown in Figs.~\ref{fig:ModTypes}(a,b). In particular, similar to the FSR-coupled device of Fig.~\ref{fig:ModTypes}(b), the proposed modulator makes use of multiple resonant modes and modulation-induced coupling between them, as shown in Figs.~\ref{fig:ModTypes}(c,d), to decouple the speed from the cavity photon lifetime. This eliminates the speed-efficiency tradeoff inherent in singly-resonant devices such as the regular microring modulator \cite{Ehrlichman2018}. Additionally, the frequency separation between the resonant modes is set by the strength of the coupling between the two cavities, as opposed to FSR-coupled devices where the resonance frequency separation is determined by the resonator radius. Therefore, in the proposed modulator, the size of the cavities can be small, which provides tight confinement of optical and RF electric fields and strong overlap between them. Additionally, the small capacitance of each active cavity permits large RC-time-limited bandwidth, which is discussed in detail in Sec.~\ref{sec:RfResAnalysis}. 

We previously proposed coupled-cavity modulators similar to Fig.~\ref{fig:GenModOptRFConf}(b) for on-chip wavelength conversion \cite{wade2014} and modulation of high-carrier-frequency RF bandlimited signals \cite{Ehrlichman2016, Ehrlichman2017, Gevorgyan2018}. The use of electro-optic coupled resonators for quantum microwave-to-optical conversion has been studied in \cite{loncar2017}. It should be noted that dual-ring modulators were also proposed prior to these works \cite{Yu2014}, but these were designed for baseband data modulation and operated in a different regime.

In the remainder of this paper we analyze in detail the different optical and electrical designs of the proposed triply-resonant modulator, illustrated in Figs.~\ref{fig:GenModOptRFConf}(b-f). We start off with a qualitative description of operation of the both optical and the RF parts of the device in Sec.~\ref{sec:OpPrinc}. In Sec.~\ref{sec:OptResAnalysis}, using the formula for conversion efficiency derived in the Appendix A, we study the basic and the generalized coupled-cavity structures from Figs.~\ref{fig:GenModOptRFConf}(b,c) and find their optimal designs that maximize the conversion efficiency. In Sec.~\ref{sec:RfResAnalysis} we explore the different RF circuits shown in Figs.~\ref{fig:GenModOptRFConf}(d-f) and derive formulae for the gain in conversion efficiency produced by the resonant circuits of Figs.~\ref{fig:GenModOptRFConf}(e) and (f) compared to the non-resonant case of Fig.~\ref{fig:GenModOptRFConf}(d). Finally, Sec.~\ref{sec:Conc} summarizes and discusses the results of this work.

\section{Principle of operation} \label{sec:OpPrinc}
Before turning to the mathematical analysis of the proposed modulator, in this section we provide a qualitative description of its optical and RF constituents and their different configurations as shown in Fig.~\ref{fig:GenModOptRFConf}.

\subsection{Problem formulation} \label{sec:formulation}
Our goal is to develop a modulator which efficiently converts RF signals into the optical domain. An RF signal modulating a continuous-wave optical input pump produces optical sidebands, and our goal is to maximize the conversion efficiency defined as the ratio of the power in the optical sideband to the input pump power, for a given RF drive power,
\begin{equation}\label{EqnConvEffDef}
	\left.\begin{aligned}
      &G=\frac{P_{sideband}}{P_{pump}}.
	\end{aligned}\right.
\end{equation}
Optimizing the modulator for the most efficient conversion into the optical sideband alone is relevant when the sideband is detected without the carrier, such as in direct-detection receivers \cite{Savchenkov2010} or photonics-assisted microwave radiometers \cite{Pett2018}. Another application is coherent communications systems, where the modulated signal is detected via the homodyne or heterodyne technique by interfering the sideband with a local oscillator.

\subsection{Optical design}
The two proposed optical modulator designs -- the basic and the generalized coupled-cavity designs -- are shown in detail in Figs.~\ref{fig:ModTypes}(c,d). The basic version of Fig.~\ref{fig:ModTypes}(c) consists of two identical evanescently coupled cavities that have the same unperturbed resonance frequency, $\omega_o$, and a bus waveguide that is coupled to one of them. The coupling strength between the resonators is described by cavity energy amplitude coupling rate $\mu$, commonly used in the coupled-mode theory (CMT) in time \cite{Little1997}. When two isolated resonators with the same resonance frequency $\omega_o$ are brought together, coupling induced splitting of supermode frequencies places the symmetric and antisymmetric supermode resonances at frequencies $\omega_o \mp \mu$, respectively, as indicated in Fig.~\ref{fig:ModTypes}(c). The coupling strength between the symmetric (antisymmetric) supermode and the input/output waveguide is characterized by the external energy amplitude decay rate $r_{e,s}$ ($r_{e,a}$). Since both supermodes are coupled to the same bus waveguide through the bottom cavity, the external energy decay rates of the supermodes are equal to each other and are equal to half of the energy amplitude coupling rate of the bottom cavity to the bus waveguide, i.e. $r_{e,s}=r_{e,a}=r_e/2$, as shown in Fig.~\ref{fig:ModTypes}(c).

If the resonance frequencies of the two cavities are modulated in time to be $\omega_o \mp \delta\omega(t)$, where the shifts $\mp\delta\omega(t)$ are induced by push-pull modulation, the instantaneous supermodes of the coupled cavity system at a given time instance are not orthogonal to the supermodes at the previous time instance. Therefore, the energy gets redistributed between the symmetric and the antisymmetric supermodes at each ``time step''. In other words, the RF drive voltage which produces the $\mp\delta\omega(t)$ resonance frequencies' modulation leads to energy coupling between the supermodes of the unperturbed system. Therefore, laser pump light at frequency $\omega_o - \mu$, entering the input port of the bus waveguide and exciting the symmetric supermode, is transferred to the antisymmetric supermode at frequency $\omega_o + \mu$ due to push-pull modulation by the RF signal at carrier frequency $\Omega_o = 2\mu$, replicating the RF spectrum in the optical sideband. The modulated light couples back into the bus waveguide and leaves through the output port.

It is essential that the resonance frequencies of the supermodes stay the same during modulation, which happens since the combined optical path length of the two microresonators does not change when their resonance frequencies are modulated in push-pull fashion. This allows the laser pump and the optical sideband to stay on resonance with the symmetric and antisymmetric modes throughout the modulation process.

The idea underlying the proposed modulator concept is that the modulation efficiency depends on the resonant enhancement of each of the interacting waves. This is confirmed by analytic derivation of the conversion efficiency in the Appendix A, which shows that the efficiency contains a product of the Lorentzian lineshape of the symmetric resonance at the frequency of the pump laser and the Lorentzian lineshape of the antisymmetric resonance at the frequency of the sideband, see Eq.~(\ref{EqnConvEffSplitResContribution}). For the moment, we only consider the efficiency limitation due to photon lifetime in the optical part of the modulator; the RF frequency response of the circuits is considered later.

The above has several implications for the modulator performance. First, for a given RF field inside the active optical cavities, efficiency of the proposed modulator does not degrade when the RF carrier frequency is much larger than the linewidth of optical resonances, much like the efficiency of the FSR-coupled modulator shown in Fig.~\ref{fig:ModTypes}(b). Second, for the best modulation efficiency, the laser pump should always be aligned to the frequency of the symmetric resonance (under the assumption that the RF carrier frequency is matched to the resonance frequency spacing). Third, when the input laser is aligned to the symmetric resonance, the shape of the small signal RF frequency response of the modulator follows the shape of the antisymmetric resonance, as can be seen from Fig.~\ref{fig:ConvEffParams}. This means that the RF bandwidth of the modulator does not depend on the linewidth of the symmetric supermode, and is equal to the linewidth of the antisymmetric supermode (which is inversely proportional to the antisymmetric mode photon lifetime). In case the RF bandwidth needs to be increased, the antisymmetric supermode linewidth needs to be broadened. In the basic design this can be achieved by increasing the coupling strength between the bus waveguide and the ring. However, this is accompanied by an unwanted reduction in Q-factor of the symmetric supermode, which reduces the pump enhancement and degrades the efficiency of the modulator.

This brings us to the generalized configuration shown in Fig.~\ref{fig:ModTypes}(d). The key idea of the generalized configuration is to enable independent control of the Q-factors of the symmetric and the antisymmetric supermodes, so that high Q-factor for the symmetric resonance can be maintained for maximum pump enhancement, while the Q-factor of the antisymmetric resonance can be reduced just enough to accommodate the modulated optical signal bandwidth within the resonance linewidth, as illustrated in Fig.~\ref{fig:ModTypes}(d). The independent control of the Q-factors of the symmetric and the antisymmetric modes is achieved with a novel interferometrically coupled input/output bus waveguide configuration, where the symmetric supermode is coupled only to the input waveguide and the antisymmetric supermode is coupled only to the output waveguide (analogous to a design for nonlinear optics \cite{Zeng2015}). In this case, the external energy decay rates of the two supermodes can also be set independently.

The interferometric coupling in the generalized coupled-cavity modulator of Fig.~\ref{fig:ModTypes}(d) works in the following way. The pump laser light, entering the input waveguide, couples into both rings in-phase, exciting the symmetric supermode. The in-phase coupling is ensured by designing the optical path difference between the two rings to be an integer of $2 \pi$. The light from the symmetric supermode can couple back to the input waveguide, but not to the output waveguide because of destructive interference of the waves coupled from each of the two rings into the output waveguide. The destructive interference is ensured by having a $\pi$ phase shifter in the output waveguide between the two rings, so that the light from the symmetric mode coupled into the output waveguide through the first and the second rings interfere destructively and cancel each other out. As a result, there is no net coupling of light from the symmetric mode to the output waveguide. The situation is reversed for the antisymmetric supermode, which has fields in the two rings oscillating out of phase with respect to each other. The waves coupled from the antisymmetric supermode into the input waveguide interfere destructively and the waves coupled into the output waveguide interfere constructively, so that the antisymmetric supermode is coupled only to the output but not to the input waveguide. The electric field configurations of the symmetric and antisymmetric supermodes are illustrated in Fig.~\ref{fig:ModTypes}(d). The energy amplitude coupling rates from the symmetric supermode to the input waveguide is $r_{e,s}=2r_{e,in}$, and the coupling rate from the antisymmetric supermode to the output waveguide is $r_{e,a}=2r_{e,out}$, where $r_{e,in}$ and $r_{e,out}$ are energy amplitude coupling rates between the individual rings and the input and output waveguides, respectively, as indicated in Fig.~\ref{fig:ModTypes}(d). By adjusting the gaps between the rings and each of these waveguides, the external decay rates $r_{e,s}$ and $r_{e,a}$ can be adjusted independently.

Note that in the special case when the external energy decay rates $r_{e,s}$ and $r_{e,a}$ of the generalized design equal those of the basic design, the two designs become equivalent and are expected to have equal modulation efficiencies (provided that other device parameters, such as losses and phase shifter efficiencies, are identical).

\subsection{RF configurations}
Different ways in which the RF drive can be applied to the active cavities of the proposed modulator are illustrated in Figs.~\ref{fig:GenModOptRFConf}(d-f). In the simplest, \textit{non-resonant} scenario, shown in Fig.~\ref{fig:GenModOptRFConf}(d), the transmission lines are directly connected to the terminals of the capacitive electro-optic region of the active cavities. From the RF perspective, the active cavity is a capacitor $C_m$ connected in series with a resistor $R_m$. The capacitor acts as a phase shifter that tunes the resonance frequency of the optical cavity in response to the RF signal, either by means of electrical charge accumulated on the capacitor plates \cite{lipson2014,watts2014} or by the electric field between them changing refractive index of an electro-optic material\cite{leuthold2017,Wang2018}. The resistor accounts for the parasitic series resistance between the capacitor plates and the terminals of the active cavities. For a fixed RF input power, the voltage on the capacitor plates and, therefore, resonance frequency modulation is maximized when the RF power is completely dissipated on the active cavities. In the non-resonant configuration of Fig.~\ref{fig:GenModOptRFConf}(d), however, part of the RF power is reflected back to the source, due to the termination of the transmission lines by the unmatched load of the active cavities. This can be considered a consequence of the frequency of the RF signal not matching the frequency of the RF resonance, which in the absence of a series inductance can be viewed as being infinite. This is illustrated in the bottom plot in Fig.~\ref{fig:GenModOptRFConf}(d), which sketches the frequency dependence of the power reflection coefficient $\left|\Gamma\right|^2$.

The RF resonance frequency can be shifted to a finite frequency by connecting an inductor in series with each active cavity. This \textit{resonant} configuration is shown in Fig.~\ref{fig:GenModOptRFConf}(e). By appropriately choosing the inductance $L_1$, the frequency of the RF resonator formed by the inductor $L_1$ and the capacitor $C_m$ can be matched to the carrier frequency of the RF drive, i.e. $1/\sqrt{L_1C_m}=\Omega_o$ [the bottom plot in Fig.~\ref{fig:GenModOptRFConf}(e)]. RF resonance boosts the voltage on the capacitors of the active cavities, improving the efficiency of the modulation. The resonance removes the reactive load from the transmission line, however, perfect load matching is still not guaranteed due to the parasitic resistance $R_m$. Therefore, the dip in reflection function, plotted in Fig.~\ref{fig:GenModOptRFConf}(e), does not reach zero, in general, even at resonance frequency.

Critical coupling between the RF feed line and the resonator can be achieved by introducing an impedance transforming circuit between them, as shown in the \textit{resonant-matched} configuration in Fig.~\ref{fig:GenModOptRFConf}(f). The figure shows an L-match impedance down-converter, which consists of capacitor $C_2$ and inductor $L_2$, and converts the higher resistance of the load to the lower characteristic impedance of the transmission line \cite{Lee2004}. At critical coupling, the dip in the reflection function reaches zero, as shown in Fig.~\ref{fig:GenModOptRFConf}(f) and field is maximized on the load capacitor, i.e. the EO region of the modulator cavities.

A detailed analysis of the different RF schemes is carried out in Sec.~\ref{sec:RfResAnalysis}, where the parasitic resistances of the inductors are taken into account. It is shown that resonant and resonant-matched circuits, implemented with CMOS inductors with typical Q-factors of $\sim$10-30 \cite{Jiang2000, Dickson2005}, provide substantial $\sim$5-20~dB gain in conversion efficiency relative to the non-resonant scheme.

\section{Analysis of the optical design} \label{sec:OptResAnalysis}

The designs of the optical and the electrical parts of the modulator can be considered independently because the electrical circuits provide the voltage which determines the optical resonant frequency swing $\delta \omega_m$, which then leads to modulation of the optical signal. This section gives a detailed analysis of the performance of the optical part of the modulator, specifically the basic and generalized configurations shown in Fig.~\ref{fig:ModTypes}(c, d), for given $\delta \omega_m$. The electrical configuration which determines this $\delta \omega_m$ is analyzed in the next section.

An analytic expression for the modulator conversion efficiency can be found by applying the CMT-in-time to the two active cavities whose frequencies are modulated in push-pull fashion as $\omega_o\pm\frac{\delta\omega_m}{2}cos(\Omega t)$, where $\Omega$ is the RF frequency. The derivation is provided in the Appendix~A. The resulting formula for the conversion efficiency $G$ which is applicable to both basic and generalized configurations is

\begin{equation}\label{EqnConvEff}
      G =\frac{\frac{1}{4}r_{e,s} r_{e,a}\delta\omega_m^2}{\left[(r_o+r_{e,a})\Delta\omega_1+(r_o+r_{e,s})\Delta\omega_2\right]^2+\left[(r_o+r_{e,s})(r_o+r_{e,a})+\left(\dfrac{\delta\omega_m}{4}\right)^2-\Delta\omega_1\Delta\omega_2\right]^2}
\end{equation}

\begin{figure}[t]
\centering
\mbox{\includegraphics{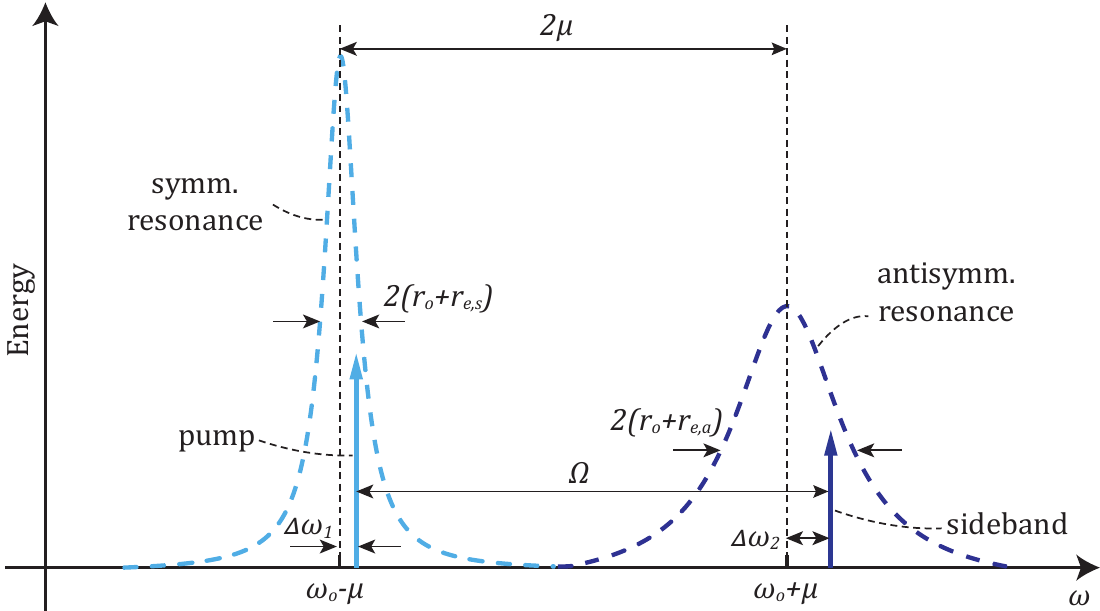}}
\captionsetup{width=.9\textwidth, font=small}
\caption{Graphical representation of the parameters in conversion efficiency formula (\ref{EqnConvEff}), with decay rates $r_o$ and $r_{e,s}$ ($r_{e,a}$) determining the linewidth of the symmetric (antisymmetric) supermode resonance, shown by light (dark) blue dashed line, coupling rate $\mu$ determining the frequency splitting between the supermode resonances, RF frequency $\Omega$ setting the separation between the laser pump (light blue arrow) and the optical sideband (dark blue arrow), and $\Delta\omega_1$ ($\Delta\omega_2$) representing the detuning of the laser pump (optical sideband) from the symmetric (antisymmetric) resonance. The relationship between $\mu$, $\Omega$, $\Delta\omega_1$ and $\Delta\omega_2$ is given by Eq.~(\ref{EqnDw2MinusDw1}).}
\label{fig:ConvEffParams}
\end{figure}

In this formula, $r_o$ is the intrinsic decay rate of the energy amplitude of the supermodes due to linear losses in cavities (which is equal to the decay rate of the modes of the individual cavities, assumed equal here), and $r_{e,s}$ and $r_{e,a}$ are the decay rates of the amplitudes of the symmetric and the antisymmetric supermodes, respectively, due to coupling to the input and output waveguides. For the basic design, which has just one waveguide coupled to a ring with coupling rate $r_e$, the supermode coupling rates are $r_{e,s} = r_{e,a} = r_e/2$ and are always equal to each other. In the generalized design, the input waveguide is coupled to each of the rings with rate $r_{e,in}$ and the output waveguide is coupled to each of the rings with rate $r_{e,out}$; in this case, the supermode coupling rates are $r_{e,s}=2r_{e,in}$ and $r_{e,a}=2r_{e,out}$, and can be chosen independently from each other. Energy amplitude coupling rate $\mu$ determines the strength of coupling between the two cavities; this coupling leads to $2 \mu$ frequency splitting of the two supermodes. The frequency separation of the laser pump from the symmetric resonance is denoted by $\Delta\omega_1$, and the frequency separation between the generated optical sideband and the antisymmetric resonance is denoted by $\Delta\omega_2$. These parameters are illustrated in Fig.~\ref{fig:ConvEffParams}. The values of $\Delta\omega_1$ and $\Delta\omega_2$ are related to $\mu$ and $\Omega$ through
\begin{equation}\label{EqnDw2MinusDw1}
	\left.\begin{aligned}
      &\Delta\omega_2-\Delta\omega_1 = \Omega-2\mu.
	\end{aligned}\right.
\end{equation}
Note that while the RF frequency $\Omega$ is not explicitly present in Eq.~(\ref{EqnConvEff}), the efficiency does depend on $\Omega$ through the above relation.

The discussion below examines the maximum conversion efficiency that can be achieved with the two proposed modulator configurations. First, we consider a single-frequency RF signal, and find the parameters which maximize the conversion efficiency at this frequency. We then find the RF bandwidth obtained in this modulator. If an RF bandwidth different from the obtained bandwidth is desired, the parameters of the modulator can be changed to ensure the required bandwidth; this is done in the last part of this section which optimizes the modulator parameters for maximum conversion efficiency while ensuring the required RF bandwidth.

First, we find the optimum conditions for modulation with an RF signal at single frequency $\Omega_o$. By inspection of Eq.~(\ref{EqnConvEff}), it is clear that the conversion efficiency is maximized when $\Delta\omega_1=\Delta\omega_2 = 0$, regardless of the values of other parameters. This is not surprising, since under this condition pump and optical sideband fields are maximally enhanced in the resonators. From Eq.~(\ref{EqnDw2MinusDw1}), it follows that the coupling rate $\mu$ between the cavities must be designed so that $\Omega_o = 2 \mu$, i.e. the frequency splitting between the resonances must match the RF frequency, in agreement with the discussion in the preceding sections. 

The next step is to find the optimum coupling strengths. The conversion efficiency is maximized for $\Omega_o =2\mu$ when
\begin{equation}\label{EqnOptimumEqualDecayRates}
	\left.\begin{aligned}
      &r_{e,s}=r_{e,a}=\sqrt{r_o^2+\left(\dfrac{\delta\omega_m}{4}\right)^2}.
	\end{aligned}\right.
\end{equation}
Here, $\delta\omega_m/4$ corresponds to the energy amplitude coupling rate between the supermodes, see Eq.~(\ref{EqnModeAmpEnvlpTilde}) in the Appendix~A. For a very weak modulating signal when the resonance frequency swing is much smaller than the intrinsic resonance linewidth, i.e. $\delta \omega_m \ll 2r_o$, this simplifies to $r_{e,s}=r_{e,a}=r_o$, which is the critical coupling condition for the supermodes. If the modulation is not weak, the above formula can be viewed as a modified critical coupling condition which takes into account the modulation-induced coupling between the supermodes, which acts as an additional source of loss for the supermodes. Note that according to Eq.~(\ref{EqnOptimumEqualDecayRates}), both of the two proposed configurations of Figs.~\ref{fig:ModTypes}(c,d) have the same supermode coupling coefficients, and show identical performance. 

\begin{figure}[t]
\centering
\mbox{\includegraphics{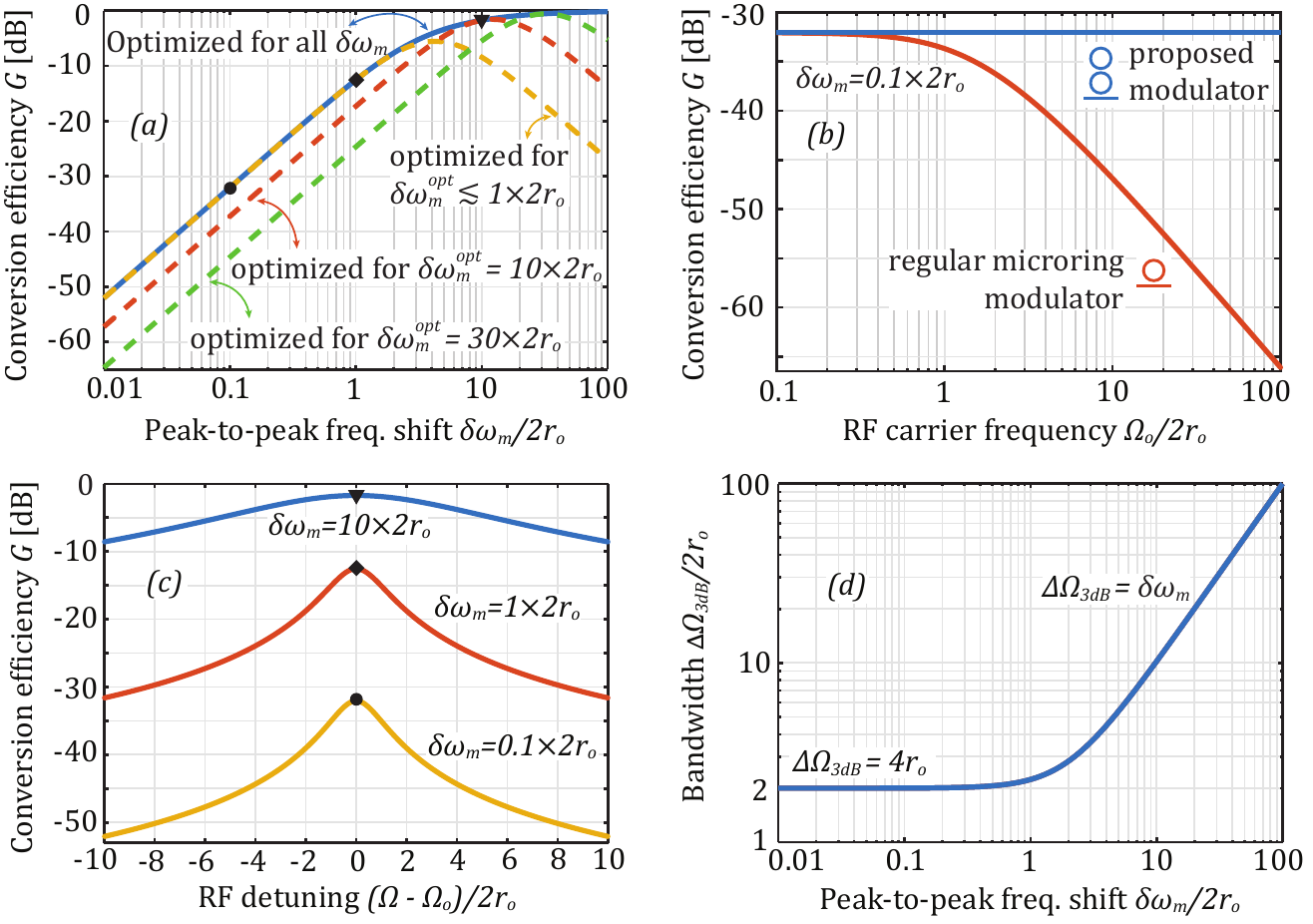}}
\captionsetup{width=.9\textwidth,font=small}
\caption{(a) Peak conversion efficiency versus resonance frequency swing $\delta\omega_m$, where $r_{e,s}$ and $r_{e,a}$ are optimized according to Eq.~(\ref{EqnOptimumEqualDecayRates}) either at each point along the x-axis (solid blue line), or for each of several values of resonance frequency swing $\delta\omega_m^{opt}$ (dashed lines). (b) Small-signal conversion efficiencies of the proposed and the regular microring modulators versus RF carrier frequency, where the modulators are optimized for maximum efficiency at each point along x-axis. (c) RF-to-optical conversion frequency response of the modulator, optimized for maximum conversion at $\Omega = \Omega_o$, for several values of $\delta\omega_m$. (d) Photon-lifetime-limited RF bandwidth, defined as full width at half maximum of the magnitude response, versus resonance frequency swing $\delta\omega_m$.}
\label{fig:ConvEffRespBw}
\end{figure}

After substitution of the optimum values for $\Delta\omega_1$, $\Delta\omega_2$, $r_{e,s}$, and $r_{e,a}$, the expression (\ref{EqnConvEff}) becomes a function of the cavity losses and the resonance frequency swing only. Fig.~\ref{fig:ConvEffRespBw}(a) plots the conversion efficiency versus normalized resonance frequency swing, i.e. the ratio of the frequency swing to the intrinsic resonance linewidth $\delta\omega_m/2r_o$. For the blue line, the modulator parameters are optimized at each point along the x-axis. When $\delta\omega_m \lesssim 2r_o$, conversion efficiency increases quadratically with the resonance frequency swing. This is the weak modulation regime where Eq.~(\ref{EqnConvEff}) reduces to
\begin{equation}\label{EqnPeakConvEffWeakSig}
	\left.\begin{aligned}
      &G(\Omega_o)=\frac{1}{4}\left(\frac{\delta\omega_m}{4}\frac{1}{r_o}\right)^2.
	\end{aligned}\right.
\end{equation}
In this regime the conversion efficiency is proportional to the square of the ratio of coupling rate between the supermodes $\delta\omega_m/4$ and the intrinsic decay rate $r_o$. This means that the efficiency of the modulator is higher if there is a higher probability of a pump photon being converted to a sideband photon, compared to the probability of losing the photon to different loss mechanisms in the cavities. When the modulation is very strong, i.e. $\delta\omega_m \gg 2r_o$, almost all the light from the pump can be converted to the sideband. The dashed lines in Fig.~\ref{fig:ConvEffRespBw}(a) show the dependence of the conversion efficiency on $\delta\omega_m$ for devices optimized for a specific resonance frequency swing $\delta\omega_m^{opt}$. If the resonance frequency swing is smaller than the one the modulator is optimized for ($\delta\omega_m \lesssim \delta\omega_m^{opt}$), the conversion efficiency increases with the swing, reaching its maximum value at $\delta\omega_m=\delta\omega_m^{opt}$. As $\delta\omega_m$ increases further, the conversion efficiency drops, which happens due to backward energy transfer from the antisymmetric to the symmetric supermode. 

It should be noted that the modulation efficiency does not depend on the frequency $\Omega_o$, if for each frequency the coupling strength between the cavities is designed so that $\Omega_o = 2\mu$ (assuming drive strength $\delta\omega_m$ does not depend on $\Omega_o$). This is in stark contrast to the modulation efficiency of the regular microring resonator modulators, which drops as the signal frequency increases. A thorough analysis of the regular microring modulators for optimal RF-to-optical conversion is given in \cite{Ehrlichman2018}. As an example, in Fig.~\ref{fig:ConvEffRespBw}(b) the small-signal conversion efficiencies of the proposed and the regular microring modulators are plotted versus the RF signal frequency normalized by the cavity resonance intrinsic linewidth $2r_o$. At each frequency $\Omega_o$ the modulators are optimized for maximum conversion efficiency. In the case of the proposed modulator it is done by adjusting the coupling strength between cavities; in the case of the microring modulator, by adjusting the cavity resonance linewidth through an appropriate choice of ring-bus coupling strength. As expected, the efficiency of the proposed modulator stays constant, while the efficiency of the microring modulator drops (20 dB/decade) as the RF frequency increases.

Let us now find the RF bandwidth of the modulator optimized for the best performance at a single RF frequency. Strictly speaking, Eq.~(\ref{EqnConvEff}) is applicable to harmonic RF input only. However, for weak modulating signal that does not deplete the laser pump, formula (\ref{EqnConvEff}) can be applied to each spectral RF component independently and used to study the frequency response of the modulator. To find the frequency response, the laser frequency is set at the resonance frequency of the symmetric supermode ($\Delta\omega_1 = 0$) and the RF frequency $\Omega$ is swept around $\Omega_o = 2\mu$, which, according to Eq.~(\ref{EqnDw2MinusDw1}), is equivalent to sweeping $\Delta\omega_2$ around zero.

The magnitude response for several values of resonance frequency swing $\delta\omega_m$ is plotted in Fig.~\ref{fig:ConvEffRespBw}(c). For each $\delta\omega_m$, the external coupling rates are found from Eq.~(\ref{EqnOptimumEqualDecayRates}). The maximum values of each response are indicated on the blue curve in Fig.~\ref{fig:ConvEffRespBw}(a) with a matching black marker. Note that with an increase in $\delta\omega_m$ not only does the peak conversion efficiency increase but the spectral response becomes wider.

The RF bandwidth of the modulator $\Delta\Omega_{3dB}$, defined as full width at half maximum of the magnitude response centered at the carrier frequency $\Omega_o$, can be found from Eq.~(\ref{EqnConvEff}) to be
\begin{equation}\label{EqnBandwidthOptimumDecayRates}
	\left.\begin{aligned}
      &\Delta\Omega_{3dB}=4\sqrt{r_o^2+\left(\dfrac{\delta\omega_m}{4}\right)^2}.
	\end{aligned}\right.
\end{equation}
Note that the RF bandwidth is simply equal to the optical bandwidth of the antisymmetric resonance, which is 4 times the critical coupling rate given by Eq.~(\ref{EqnOptimumEqualDecayRates}). Figure~\ref{fig:ConvEffRespBw}(d) plots the modulation bandwidth versus peak-to-peak resonance frequency swing $\delta\omega_m$. When $\delta\omega_m \lesssim 2r_o$, the bandwidth is limited to twice the intrinsic linewidth of the ring cavities. As modulation becomes stronger and $\delta\omega_m$ increases, the effective energy escape rate in the supermodes goes up, broadening the optical resonances and the RF bandwidth. 

The analysis above found a modulator design which provides the maximum conversion efficiency at a given RF carrier frequency and determined the resulting RF bandwidth for this design. This approach works well for narrowband RF signals, however, many applications require RF bandwidths wider than provided by the above design. Our goal now is to obtain a design which not only maximizes the efficiency at given carrier frequency $\Omega_o$, but also meets a minimum required RF bandwidth.

Expression (\ref{EqnBandwidthOptimumDecayRates}) is only valid when the external coupling rates are given by (\ref{EqnOptimumEqualDecayRates}). A more general expression for the 3dB bandwidth with arbitrary, independent $r_{e,s}$ and $r_{e,a}$ is 
\begin{equation}\label{EqnBandwidthEqualDecayRates}
	\left.\begin{aligned}
      &\Delta\Omega_{3dB}=2(r_o+r_{e,a}) + \dfrac{1}{2(r_o+r_{e,s})}\left(\dfrac{\delta\omega_m}{2}\right)^2.
	\end{aligned}\right.
\end{equation}
\begin{figure}[!t]
\centering
\mbox{\includegraphics{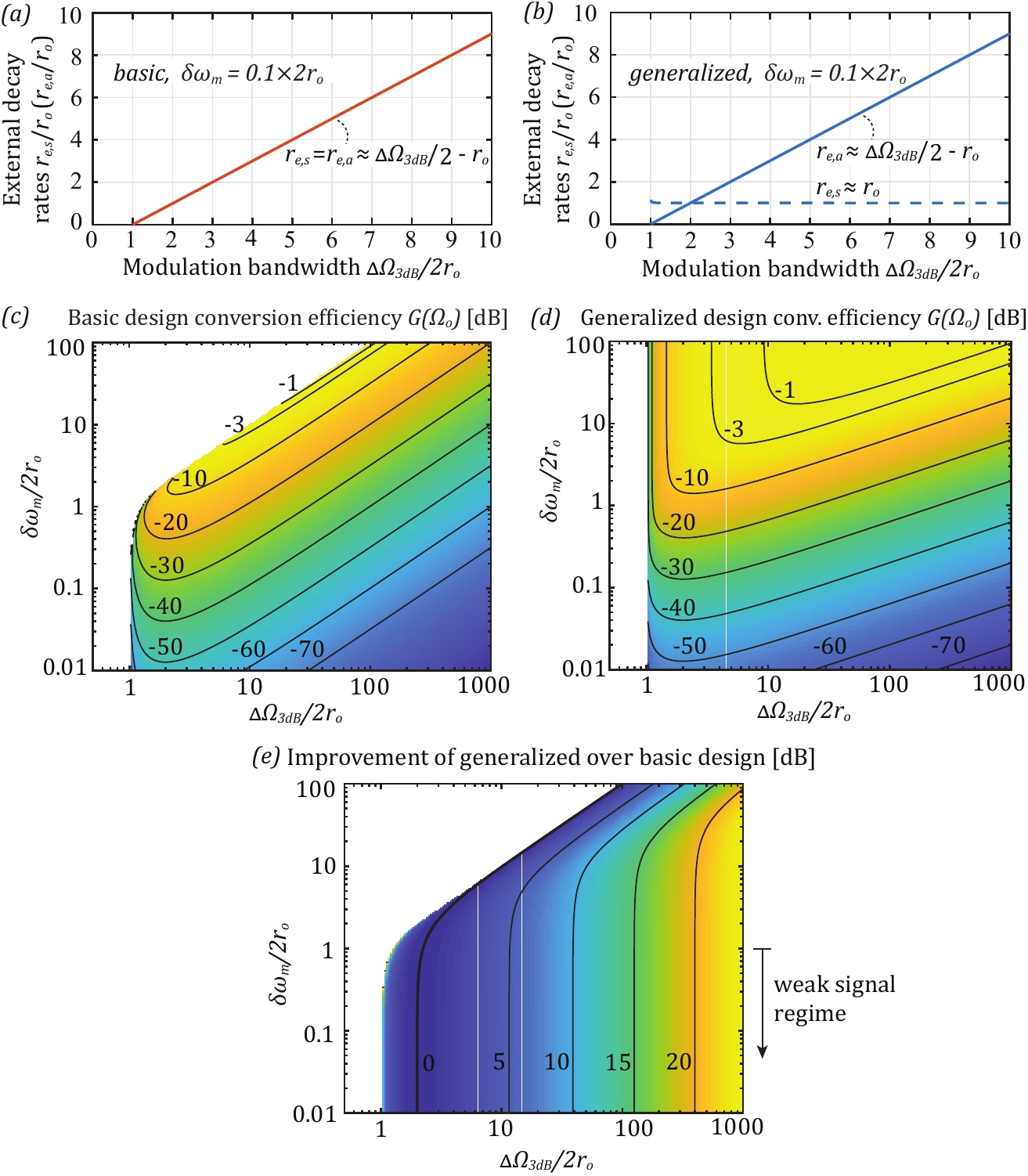}}
\captionsetup{width=.9\textwidth,font=small}
\caption{(a,b) Design external decay rates $r_{e,s}$ and $r_{e,a}$ versus RF bandwidth $\Delta\Omega_{3dB}$ for (a) the basic and (b) the generalized designs, optimized for maximum conversion efficiency. (c,d) Peak conversion efficiency versus $\Delta\Omega_{3dB}$ and $\delta\omega_m$ for (c) the basic and (d) the generalized designs. (e) The improvement in conversion efficiency that the generalized design provides over the basic design versus $\Delta\Omega_{3dB}$ and $\delta\omega_m$.}
\label{fig:ConvEffVsBw}
\end{figure}
In the above expression, the second term can usually be neglected in comparison to the first term except when the modulation is strong. Therefore, to increase the RF bandwidth, one needs to increase $r_{e,a}$, the coupling rate of the antisymmetric mode (only), increasing its optical bandwidth. Widening of the bandwidth comes at the expense of reduced conversion efficiency due to an increase of the denominator in Eq.~(\ref{EqnConvEff}). In the basic design, $r_{e,a}$ can be increased by increasing the coupling between the microresonator and the bus waveguide. However, as discussed earlier, the coupling rate $r_{e,s}$ increases by the same amount, which causes an added reduction in conversion efficiency according to Eq.~(\ref{EqnConvEff}). In contrast, in the generalized architecture of Fig.~\ref{fig:ModTypes}(d), $r_{e,a}$ can be increased by increasing the coupling of the rings to the output waveguide, without changing the coupling to the input waveguide and changing $r_{e,s}$. Thus, broadband modulation can be realized with significantly higher efficiency in the generalized than in the basic design, in principle. 

To quantify the improvement that the generalized design provides over the basic design, the conversion efficiency formula (\ref{EqnConvEff}) must be expressed in terms of $\Delta\Omega_{3dB}$. This is straightforward for the basic architecture. The external coupling rates are set to be equal in Eq.~(\ref{EqnBandwidthEqualDecayRates}) and are expressed in terms of $\Delta\Omega_{3dB}$
\begin{equation}\label{EqnEqualDecayRatesInTermsOfBandwidth}
	\left.\begin{aligned}
      &r_{e,s} = r_{e,a}=\dfrac{\Delta\Omega_{3dB}}{4}-r_o + \sqrt{\left(\dfrac{\Delta\Omega_{3dB}}{4}\right)^2 - \left(\dfrac{\delta\omega_m}{4}\right)^2}, \\[6pt] 
      &\mbox{and}\quad\Delta\Omega_{3dB}^{min}=
      \begin{cases}
      2r_o+\dfrac{1}{2r_o}\left(\dfrac{\delta\omega_m}{2}\right)^2, & \mbox{if } \delta\omega_m \leq 4 r_o \\ \delta\omega_m\ , & \mbox{if } \delta\omega_m > 4 r_o
      \end{cases}
	\end{aligned}\right.
\end{equation}
is the lower limit of the RF bandwidth. Finally, the expression for $r_{e,s}$ and $r_{e,a}$ from Eq.~(\ref{EqnEqualDecayRatesInTermsOfBandwidth}) are substituted into Eq.~(\ref{EqnConvEff}). For the generalized architecture the process is slightly more involved. One of the decay rates is found from Eq.~(\ref{EqnBandwidthEqualDecayRates}), substituted into Eq.~(\ref{EqnConvEff}) and the optimum value of the second decay rate is found among the extrema of Eq.~(\ref{EqnConvEff}), which results in

\begin{equation}\label{EqnUnequalDecayRatesInTermsOfBandwidth}
	\begin{aligned}
      &r_{e,s}=\frac{1}{8(\Delta\Omega_{3dB}-2r_o)}\left[\delta\omega_m^2 + \sqrt{\left[\delta\omega_m^2-8r_o(\Delta\Omega_{3dB}-2r_o)\right]^2 + 8r_o\delta\omega_m^2(\Delta\Omega_{3dB}-2r_o)}\right],\\[6pt]
      &r_{e,a}=\frac{\Delta\Omega_{3dB}-2r_o}{3} - \frac{1}{48r_o}\left[\delta\omega_m^2 - \sqrt{\left[\delta\omega_m^2-8r_o(\Delta\Omega_{3dB}-2r_o)\right]^2 + 8r_o\delta\omega_m^2(\Delta\Omega_{3dB}-2r_o)}\right], \\[6pt]
      &\mbox{and}\quad \Delta\Omega_{3dB}^{min}=2r_o.
      \end{aligned}
\end{equation}
Finally, the above expressions for coupling rates are substituted into Eq.~(\ref{EqnConvEff}). In the limit when the modulation is weak ($\delta\omega_m \lesssim 2r_o$) and the required photon-lifetime-limited bandwidth is much larger than the intrinsic linewidth ($\Delta\Omega_{3dB} \gg 2r_o$), Eqs.~(\ref{EqnEqualDecayRatesInTermsOfBandwidth}) and (\ref{EqnUnequalDecayRatesInTermsOfBandwidth}) simplify to

\begin{equation}\label{EqnModeAmpEnvlp1}
	\left.\begin{aligned}
      &r_{e,s} = r_{e,a} \approx \frac{1}{2}\Delta\Omega_{3dB}-r_o\quad\mbox{(basic)}, \\[6pt]
      &r_{e,s} \approx r_o, r_{e,a} \approx \frac{1}{2}\Delta\Omega_{3dB}-r_o\quad\mbox{(generalized)}.
	\end{aligned}\right.
\end{equation}

In Figs.~\ref{fig:ConvEffVsBw}(a) and (b) the external decay rates, given by Eqs.~(\ref{EqnEqualDecayRatesInTermsOfBandwidth}) and (\ref{EqnUnequalDecayRatesInTermsOfBandwidth}), are plotted versus the normalized bandwidth for $\delta\omega_m = 0.1\times 2r_o$. Fig.~\ref{fig:ConvEffVsBw}(b) shows that for given bandwidth, maximum conversion efficiency is achieved when the symmetric resonance is critically coupled to the input port, while the coupling strength between the antisymmetric resonance and the output port is increased until the required bandwidth is reached.

The peak conversion efficiencies of the basic and the generalized designs are plotted versus the target RF bandwidth $\Delta \Omega_{3dB}$ and the resonance frequency swing $\delta\omega_m$ in Figs.~\ref{fig:ConvEffVsBw}(c) and (d). The white regions in the plots indicate that no design exists for $\Delta\Omega_{3dB}<\Delta\Omega_{3dB}^{min}$, where $\Delta\Omega_{3dB}^{min}$ for the basic and generalized designs is given in Eqs.~(\ref{EqnEqualDecayRatesInTermsOfBandwidth}) and (\ref{EqnUnequalDecayRatesInTermsOfBandwidth}), respectively. The peak conversion efficiency of both designs drops as the target bandwidth increases, however, the drop is significantly slower in the generalized (10 dB/decade) than in the basic design (20 dB/decade). The improvement the generalized design provides over the basic design is shown in Fig.~\ref{fig:ConvEffVsBw}(e). It is negligible when the target RF bandwidth is close to the intrinsic resonance linewidth, which means that the generalized design gives little advantage over the simpler basic design if the target RF bandwidth is low, or the microrings are lossy. The advantage of the generalized design increases as the target RF bandwidth increases relative to the intrinsic linewidth of the resonators. For instance, when the required RF bandwidth is about 40 times larger than the intrinsic linewidth, the generalized design provides 10~dB higher conversion efficiency than the basic design. Therefore, generalized configuration, based on tailoring of resonance photon lifetimes, is expected to be particularly useful when a low-loss optical phase shifting mechanism is available, such as in \cite{Alexander2017,Wang2018}.

\section{Analysis of the RF design} \label{sec:RfResAnalysis}
\begin{figure}[t]
\centering
\mbox{\includegraphics{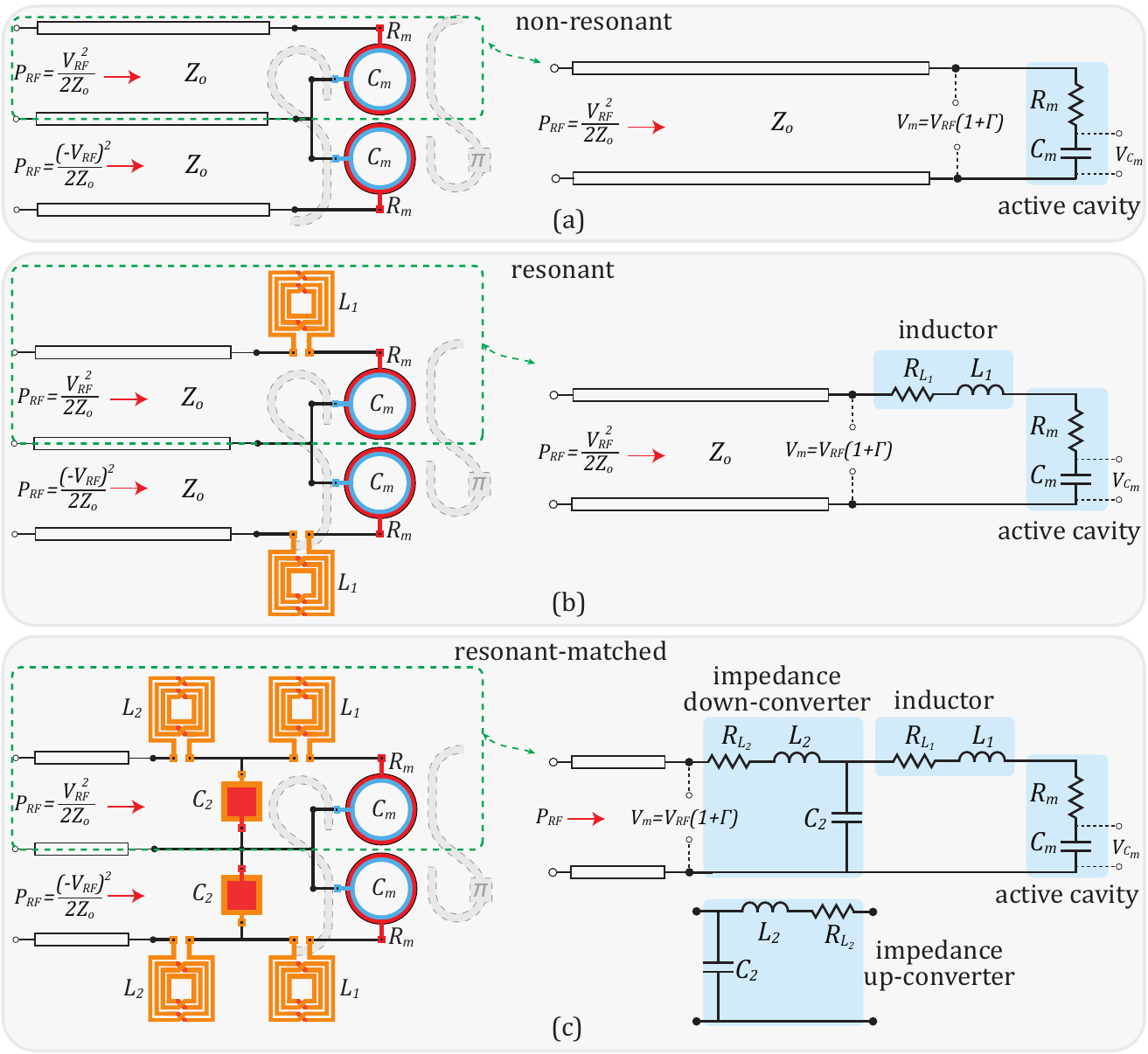}}
\captionsetup{width=.9\textwidth,font=small}
\caption{RF configurations of the proposed modulator with (a) non-resonant, (b) resonant, (c) resonant-matched designs on the left and corresponding equivalent circuits on the right.}
\label{fig:RfConfigEquivCirc}
\end{figure}

In the previous section, the analysis was focused on determining the optimum optical design of the proposed triply resonant modulator. In this section we turn our attention to the RF design, and find the gain in conversion efficiency that the \textit{resonant} and \textit{resonant-matched} circuits in Figs.~\ref{fig:GenModOptRFConf}(e) and (f) produce compared to the \textit{non-resonant} circuit in Fig.~\ref{fig:GenModOptRFConf}(d), for a given input RF power. The RF designs are replicated on the left side of Fig.~\ref{fig:RfConfigEquivCirc}, with corresponding equivalent circuits shown on the right side.

The relationship between the resonance frequency swing $\delta\omega_m$ and the input RF power $P_{RF}$ needs to be determined first. The RF signal is brought to the modulator by a transmission line with characteristic impedance $Z_o$. The voltage amplitude of the forward-propagating wave is equal to $V_{RF} = \sqrt{2Z_oP_{RF}}$. The voltage amplitude across the capacitor of the active cavity $V_{C_m}$ is equal to the product of $V_{RF}$ and the voltage enhancement frequency response of the RF circuit, $V_{C_m} = \left\vert H(\Omega)\right\vert V_{RF}$. Finally, the resonance frequency swing is related to the voltage amplitude across the capacitor of the active cavity $V_{C_m}$ through
\begin{equation}\label{EqnFreqSwingVoltAmp}
	\left.\begin{aligned}
      &\delta\omega_m = \frac{\pi c V_{C_m}}{n_g V_{\pi}L} = \frac{\pi c}{n_g V_{\pi}L}\left\vert H(\Omega)\right\vert\sqrt{2Z_o P_{RF}} \ ,
	\end{aligned}\right.
\end{equation}
where $c$ is the speed of light in vacuum, $n_g$ is the group index of the waveguide that makes up the cavities and, finally, $V_{\pi}L$ [Vm] is the voltage-length product of the capacitive phase shifter (L is typically the ring roundtrip length). According to Eq.~(\ref{EqnPeakConvEffWeakSig}), the conversion efficiency changes quadratically with $\delta\omega_m$ in the weak modulation regime. Therefore, the gain in conversion efficiency produced by the resonant and resonant-matched circuits is equal to the magnitude squared of the voltage enhancement produced by these circuits relative to the non-resonant design. 

Next, the voltage enhancement frequency responses of the different circuit configurations are found. The non-resonant circuit configuration is shown in Fig.~\ref{fig:RfConfigEquivCirc}(a). Here the two active cavities of the modulator are directly connected to the terminals of the transmission lines that deliver the differential RF signal from a signal source to the modulator. The equivalent circuit of one of the branches is shown on the right of Fig.~\ref{fig:RfConfigEquivCirc}(a), where the active cavity is represented by the series connection of the capacitor $C_m$ and the parasitic resistance $R_m$. The voltage on the terminals of the active cavity is equal to the sum of voltages of the incident and the reflected waves $V_m = V_{RF}(1+\Gamma)$, where the reflection coefficient $\Gamma$ is given by the well known expression $\Gamma = \frac{Z_m - Z_o}{Z_m + Z_o}$, and $Z_m = R_m + \frac{1}{j\Omega C_m}$ is the total impedance of the active cavity with capacitance $C_m$ and resistance $R_m$. The voltage on the capacitor of the active cavity is $V_{C_m} = \frac{1}{j\Omega C_m Z_m}V_m$, and the voltage response of the circuit is readily expressed as 
\begin{equation}\label{EqnResponseRC}
	\left.\begin{aligned}
      &H_{RC}(\Omega) \equiv \frac{V_{C_m}}{V_{RF}} = \frac{2}{1+j\Omega(Z_o + R_m)C_m}.
	\end{aligned}\right.
\end{equation}

\begin{figure}[t]
\centering
\mbox{\includegraphics{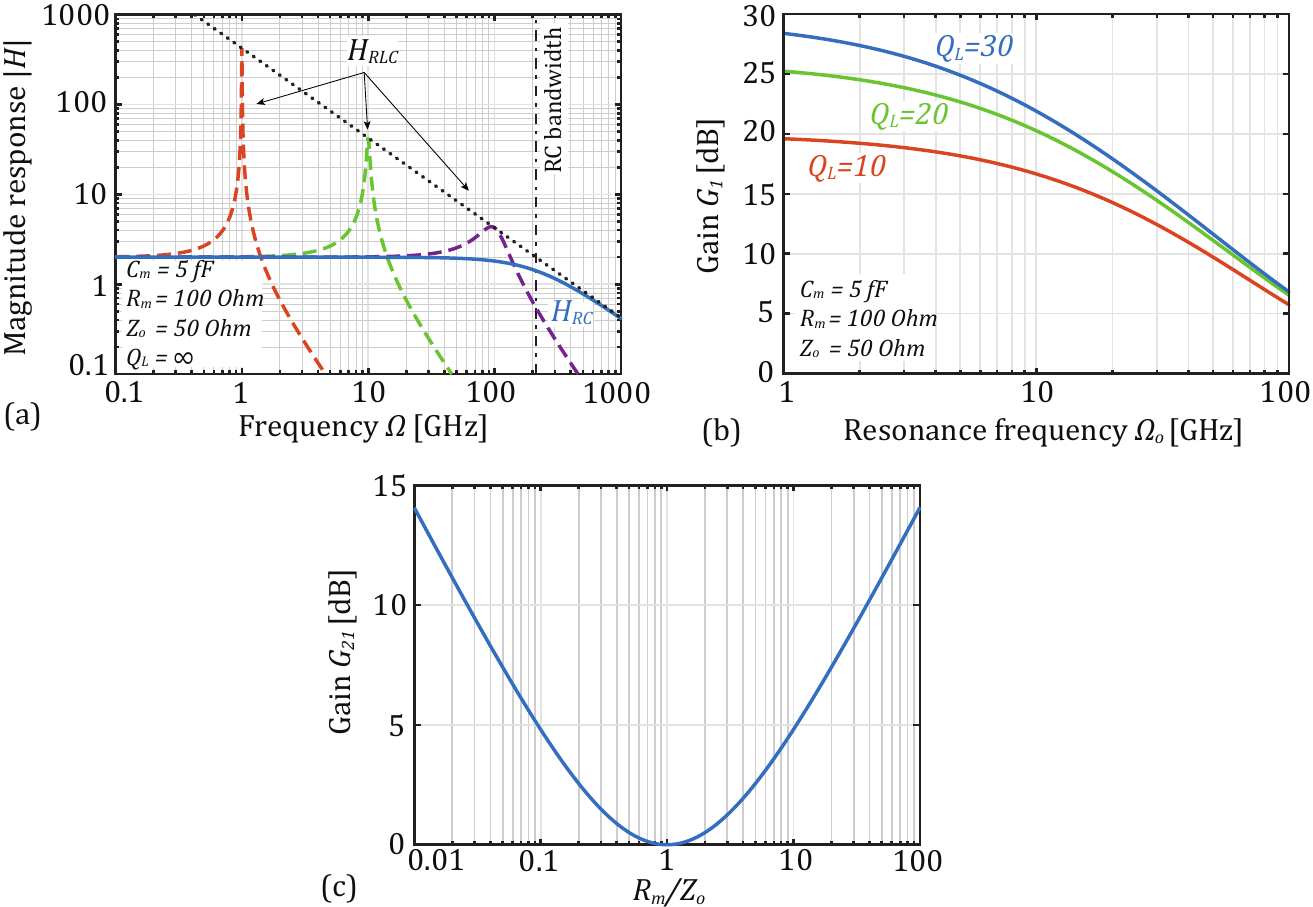}}
\captionsetup{width=.9\textwidth,font=small}
\caption{(a) Magnitude response of the non-resonant circuit of Fig.~\ref{fig:RfConfigEquivCirc}(a) (solid blue line) and of the resonant circuit of Fig.~\ref{fig:RfConfigEquivCirc}(b) for RF resonance frequency $\Omega_o$ equal to 1, 10, and 100 GHz (dashed lines). (b) The gain in conversion efficiency the resonant design provides compared to the non-resonant design versus $\Omega_o$ for several values of $Q_L$ of the integrated inductor. Values of $C_m$=5 fF, $R_m$=100 Ohm, $Z_o$=50 Ohm are assumed for the plots in (a) and (b). (c) The additional gain in conversion efficiency the resonant-matched circuit provides compared to resonant design versus normalized parasitic resistance of the active cavities.}
\label{fig:RfResGain}
\end{figure}

The factor of two in the numerator indicates that the low-frequency components reflect entirely due to the large impedance of the capacitor, doubling the voltage on the load. Note that RC-time-limited bandwidth depends not only on the parasitic resistance $R_m$, but on the impedance of the transmission line as well. The capacitance of state-of-the-art silicon microring modulators is on the order of femtofarads or several tens of femtofarads, and the resistance is between tens and hundreds of Ohms \cite{watts2014,campenhout2015}. As an example, the blue line in Fig.~\ref{fig:RfResGain}(a) shows the magnitude response of the non-resonant circuit with $C_m$=5 fF, $R_m$=100 Ohm, and a transmission line with $Z_o$=50 Ohm.

In the resonant configuration of Fig.~\ref{fig:RfConfigEquivCirc}(b), series RLC circuits are formed by the active cavities and integrated inductors. The inductance $L_1$ is chosen such that the resonance frequency of the RLC circuit is equal to the RF carrier frequency, $1/\sqrt{L_1C_m}=\Omega_o$. The equivalent circuit is shown on the right of Fig.~\ref{fig:RfConfigEquivCirc}(b). Here, the parasitic resistance of the inductor is expressed through $R_{L_1}=\Omega_oL_1/Q_L$, where $Q_L$ is the quality factor of the inductor. The frequency response of this circuit, calculated following the same steps as in the previous case is 
\begin{equation}\label{EqnResponseRLC}
	\left.\begin{aligned}
      &H_{RLC} = \frac{V_{C_m}}{V_s} = \frac{2}{1- \Omega^2L_1C_m+j\Omega (Z_o + R_m + R_{L_1})C_m}.
	\end{aligned}\right.
\end{equation}
At resonance, the real part of the denominator is zero, and the imaginary part is the inverse of the total quality factor of the RLC circuit, $Q_{RF}^{tot}=1/(Z_o + R_m + R_{L_1})\Omega_oC_m$. This means that the voltage across the capacitor is $Q_{RF}^{tot}$ times larger at resonance than at low frequencies. The magnitude of Eq.~(\ref{EqnResponseRLC}) is plotted in Fig.~\ref{fig:RfResGain}(a) [dashed lines] for three different values of the resonance frequency $\Omega_o$, assuming the same values of $C_m$, $R_m$ and $Z_o$ as in the previous example and infinite $Q_L$. As the resonance frequency approaches the RC-time-limited bandwidth, the maximum voltage gain, shown by the dotted line in Fig.~(\ref{fig:RfResGain}), diminishes, which happens due to the reduction of total quality factor of the RLC resonator.

The gain in the modulator conversion efficiency at frequency $\Omega_o$ the resonant circuit produces compared to the non-resonant circuit, is equal to the square of the voltage gain, which can be found from Eqs.~(\ref{EqnResponseRLC}) and (\ref{EqnResponseRC}):
\begin{equation}\label{EqnGainRLC}
	\left.\begin{aligned}
      &G_1(\Omega_o) = \left\lvert\frac{H_{RLC}}{H_{RC}}\right\rvert_{\Omega_o}^2 = \left(Q_{RF}^{tot}\right)^2\left[1+\left(\frac{1}{Q_{RF}^{tot}}-\frac{1}{Q_{L}}\right)^2\right].
	\end{aligned}\right.
\end{equation}
For $Q_L \gg Q_{RF}^{tot}$, this simplifies to $G_1=1+\left(Q_{RF}^{tot}\right)^2$. The dependence of $G_1$ on the resonance frequency $\Omega_o$ is shown in Fig.~\ref{fig:RfResGain}(b) for several values of $Q_L$ and the same values of $C_m$, $R_m$ and $Z_o$ as in the previous examples. For instance, at 50 GHz a resonant circuit with an inductor that has Q-factor $Q_L$=10 provides 10 dB gain. This illustrates that large gain is expected at microwave frequencies from a resonant circuit with $Q_L's$ typical of CMOS inductors \cite{Jiang2000, Dickson2005}. Even larger gains are possible for smaller capacitances and resistances of the active cavities.

It is worth noting that the RF resonance quality factor $Q_{RF}^{tot}$, as well as the optical photon-lifetime (discussed in Sec.~\ref{sec:OptResAnalysis}) determine the overall RF bandwidth of the modulator.

One can show that, in the resonant circuit of Fig.~\ref{fig:RfConfigEquivCirc}(b), the voltage $V_{C_m}$ on resonance (and for a fixed $P_{RF}$ power) is maximized if the characteristic impedance $Z_o$ matches the load resistance $R_m+R_{L_1}$. If the characteristic impedance differs from the load resistance, an impedance matching circuit can be implemented with lumped elements \cite{Lee2004}, as is done in the resonant-matched design in Fig.~\ref{fig:RfConfigEquivCirc}(c). If the load resistance is higher than $Z_o$, an L-match impedance transformer, consisting of the capacitor $C_2$ and the inductor $L_2$, can be used to down-convert the active load resistance to match it with the characteristic impedance $Z_o$, as shown in Fig.~\ref{fig:RfConfigEquivCirc}(c). If the load resistance is lower than $Z_o$ an impedance up-converter, shown on the bottom right in Fig.~\ref{fig:RfConfigEquivCirc}(c) can be used. In both cases the gain in conversion efficiency compared to the non-resonant circuit is
\begin{equation}\label{EqnResponseRLCmatched}
	\left.\begin{aligned}
      &G_2(\Omega_o) = \left\lvert\frac{H_{RLC}^{match}}{H_{RC}}\right\rvert_{\Omega_o}^2 = \frac{1}{4}\left(\sqrt{\frac{R_m}{Z_o}}+\sqrt{\frac{Z_o}{R_m}}\right)^2G_1 = G_{21}G_1,
	\end{aligned}\right.
\end{equation}
where the gain term $G_{21}$ represents the improvement over the resonant design. Here, for the sake of simplicity, we assumed the inductors have infinite quality factors. Figure~\ref{fig:RfResGain}(c) shows the dependence of $G_{21}$ on the $R_m/Z_o$ ratio. The gain that the resonant-matched design provides compared to the resonant design is about 0.5 dB when $R_m = 2Z_o$ and reaches 10 dB only when the load resistance exceeds the characteristic impedance 40 times. Since the matching circuit adds additional complexity to the system, its implementation is justified only when there is a large impedance mismatch between the load and the transmission line.

\section{Discussion and summary} \label{sec:Conc}
In this work we proposed and studied a triply-resonant modulator architecture for high-carrier-frequency RF modulation. The device, supports two optical and one RF resonances and simultaneously enhances all three interacting waves -- the laser pump, the RF drive, and the generated optical sideband -- maximizing the modulation efficiency.

The modulator was studied by analyzing the designs of its optical and RF parts separately, with Eqs.~(\ref{EqnConvEff}) and (\ref{EqnFreqSwingVoltAmp}) describing the performance of the optical and the electrical constituents. By combining these equations, the full electro-optic response of the modulator was found. 

On the optical side, it was shown that the performance of the proposed modulator is intrinsically insensitive to RF carrier frequency (as long as the modulator is designed for this frequency). This is in contrast to regular resonant modulators, where the photon lifetime degrades the performance as RF carrier frequency goes up. For the proposed modulator, the photon lifetime is decoupled from the RF carrier frequency, and determines the RF bandwidth around the carrier frequency. Two optical designs -- the basic design shown in Fig.~\ref{fig:GenModOptRFConf}(b) and the generalized design shown in Fig.~\ref{fig:GenModOptRFConf}(c) -- were considered. If the spectral width of the RF signal is lower than or on the order of the intrinsic linewidth, the basic optical design shown in Fig.~\ref{fig:GenModOptRFConf}(b) works well, and the more complex generalized design brings no performance advantages. However, the generalized design is preferable for RF signals with spectrum significantly wider than the intrinsic linewidth of the cavity. By proper engineering of loaded Q-factors of the supermode resonances in the generalized design, the conversion efficiency of wide-spectrum RF signals can be significantly increased in comparison to the basic design.

On the RF side, it was shown that the efficiency of the modulator can be enhanced by connecting inductors in series with the capacitors of the active cavities to form LC resonators with resonance frequency equal to the RF carrier frequency. The LC resonance improves the conversion efficiency relative to the case when the active cavities are connected directly to the transmission line by approximately $\left(Q_{RF}^{tot}\right)^2+1$, where $Q_{RF}^{tot}$ is the quality factor of the resonant circuit. Additionally, an impedance matching scheme was proposed for situations when there is large impedance mismatch between the RF feed line and the modulator. In the impedance-matched design, input RF power is directed to maximally build up energy (and thus voltage drop) in the active region capacitance of the modulator, maximizing its efficiency. 

The analysis of the previous sections showed that the modulator efficiency depends on a number of factors such as efficiency of the electro-optic phase shifters, optical loss in the cavities, the capacitance and the resistance of the active cavities, etc. A useful figure of merit can be found from the weak signal peak conversion efficiency formula (\ref{EqnPeakConvEffWeakSig}). Combining Eqs.~(\ref{EqnPeakConvEffWeakSig}), (\ref{EqnFreqSwingVoltAmp}), (\ref{EqnResponseRLC}), and the first expression of Eq.~(\ref{EqnCavityParamConversion}), we arrive at
\begin{equation}\label{EqnFigureOfMerit}
	\left.\begin{aligned}
      &G(\Omega_o) = 2Z_oP_{RF}\left(\frac{5\pi}{ln(10)}\right)^2\left(\frac{Q_{RF}^{tot}}{V_{\pi}L\alpha}\right)^2
	\end{aligned}\right.
\end{equation}
where $\alpha$ [dB/m] is waveguide propagation loss of the cavities. The last term in Eq.~(\ref{EqnFigureOfMerit}), which depends on $V_{\pi}L\alpha$-product of the phase shifter and total quality factor $Q_{RF}^{tot}$ of the RF resonator, can serve as a figure of merit (FOM) for the proposed modulator and can be used to compare performance of the modulator achievable in different material platforms,
\begin{equation}\label{EqnFOM}
	\left.\begin{aligned}
      &FOM = \frac{Q_{RF}^{tot}}{V_{\pi}L\alpha}.
	\end{aligned}\right.
\end{equation}

Electro-optic phase shifters with very low $V_{\pi}L\alpha$-products have been demonstrated in platforms such as lead zirconate titanate-on-silicon nitride with $V_{\pi}L$=1.02~Vcm and $V_{\pi}L\alpha$<1~VdB \cite{Alexander2017}, hybrid barium titanate-silicon with $V_{\pi}L$=0.3~Vcm and $V_{\pi}L\alpha$=1.7~VdB \cite{Eltes2017}, and silicon-organic hybrid with $V_{\pi}L$=0.032~Vcm and $V_{\pi}L\alpha$=1.2~VdB \cite{Kieninger2018}. Carrier-depletion-based electro-optic phase shifters in silicon have been demonstrated with $V_{\pi}L\alpha$-product as low as 5.7~VdB and $V_{\pi}L$=0.46~Vcm \cite{Yong2017}. Although the $V_{\pi}L\alpha$ metric for silicon is not as low as for some other platforms, silicon modulators can be readily implemented in CMOS processes (including processes with $f_T$'s up to 305 GHz \cite{45rfsoi2017} and 485 GHz \cite{Lee2007}) alongside CMOS electronics, enabling complex CMOS RF-photonic systems. Moreover, RF resonators, implemented monolithically with CMOS inductors available in these processes \cite{45rfsoi2017}, can significantly improve the modulator performance at low cost and with low parasitics.

With the increasing importance of high bandwidth RF signal processing, and the complex photonic and electronic-photonic integrated circuit platforms that are emerging \cite{sun2015}, modulators such as those proposed may find a natural place in MWP integrated circuits. In addition to experimental validation, work remains to address the linearity and power handling of these designs. Furthermore, while this paper addresses single sideband generation, the presented ideas can be applied to e.g. triple-optical-cavity systems \cite{wade2014} for dual sideband generation.

\section*{Appendix A: Derivation of the conversion efficiency formula}
\label{appndxA}
\setcounter{equation}{0}
\renewcommand{\theequation}{A.\arabic{equation}}
Here, we derive the conversion efficiency formula (\ref{EqnConvEff}), using the coupled-mode theory (CMT) in time \cite{Little1997}. The analysis below is general and applicable to both the basic and the generalized architectures of Figs.~\ref{fig:ModTypes}(c,d).

We start with the system of two coupled cavities, which are not coupled to external waveguides, and which have their resonance frequencies modulated in push-pull mode. The CMT equations for such a system can be formulated as
\begin{equation}\label{EqnCmtRingAmp}
	\left.\begin{aligned}
      &\frac{d}{dt} \bar{a}=j\bar{\bar{\omega}}\cdot\bar{a}-j\bar{\bar{\mu}}\cdot\bar{a} \\[6pt]
	\end{aligned}\right.
\end{equation}
with 
\begin{equation*}
	\left.\begin{aligned}
      &\bar{a} = 
      \begin{pmatrix}
          a_1 \\[4pt]
          a_2
      \end{pmatrix},
      \quad
      \bar{\bar{\omega}} = 
      \begin{pmatrix}
          \omega_o-\delta\omega(t)+jr_o & 0\\[4pt]
          0 & \omega_o+\delta\omega(t)+jr_o
      \end{pmatrix},
      \quad
      \bar{\bar{\mu}} = 
      \begin{pmatrix}
          0 & \mu \\[4pt]
          \mu & 0
      \end{pmatrix}.
    \end{aligned}\right.
\end{equation*}
Here, $a_1(t)$ and $a_2(t)$ are energy amplitudes of the optical fields in the microresonators which oscillate at resonance frequency $\omega_o$ in absence of modulation. The term $\delta\omega(t)$ represents the instantaneous changes in the resonance frequencies due to modulation, and different signs of $\delta\omega(t)$ in diagonal terms of the $\bar{\bar{\omega}}$ matrix indicate that the resonance frequencies of the two rings are modulated in push-pull mode. Energy amplitude decay rate $r_o$ accounts for the intrinsic losses in the microresonators, and is assumed equal in both here. The derivations are easily modified to remove this assumption. 

By solving Eq.~(\ref{EqnCmtRingAmp}) in the absence of the modulating signal ($\delta\omega(t) = 0$), we find the eigenvectors (supermode fields) and the eigenvalues (resonant frequencies) of the coupled-cavity system:
\begin{equation}\label{EqnEigenstates}
	\left.\begin{aligned}
      &\begin{pmatrix}
          a_1 \\[4pt]
          a_2
      \end{pmatrix}_{s,a}=
      \frac{1}{\sqrt{2}}
      \begin{pmatrix}
          1 \\[4pt]
          \pm1
      \end{pmatrix}, 
	\end{aligned}\right.
\end{equation} 
\begin{equation}\label{EqnEigenvalues}
	\left.\begin{aligned}
      &\omega_{s,a} = \omega_o \mp \mu + jr_o.
      \end{aligned}\right.
\end{equation}
The two supermodes are referred to as ``symmetric'' and ``antisymmetric'' because according to Eq.~(\ref{EqnEigenstates}), the fields in the two rings are in phase for one supermode and out of phase for the other. According to Eq.~(\ref{EqnEigenvalues}), resonance frequencies of the supermodes, given by the real parts of the eigenvalues $\omega_{s,a}$, are split by twice the energy amplitude coupling rate $\mu$ between the rings. Additionally, the decay rate of each supermode due to losses in the cavities, given by the imaginary parts of $\omega_{s,a}$, is the same as cavity intrinsic decay rate $r_o$. 

The CMT equations (\ref{EqnCmtRingAmp}) can be rewritten in terms of the supermodes of the unperturbed system (i.e. the system in absence of modulation),
\begin{equation}\label{EqnModeAmp}
	\left.\begin{aligned}
      &\frac{d}{dt} \bar{b}=j\bar{\bar{\omega}}\cdot\bar{b}-j\bar{\bar{\mu}}\cdot\bar{b}
	\end{aligned}\right.
\end{equation}
where $b_1$ and $b_2$ are the energy amplitudes of the supermodes,
\begin{equation*}
	\left.\begin{aligned}
      &\bar{b} = 
      \begin{pmatrix}
          b_1 \\[4pt]
          b_2
      \end{pmatrix}=\frac{1}{\sqrt{2}}
      \begin{pmatrix}
          a_1+a_2 \\[4pt]
          a_1-a_2
      \end{pmatrix},
          \end{aligned}\right.
\end{equation*}
and
\begin{equation*}
	\left.\begin{aligned}
      \quad
      &\bar{\bar{\omega}} = 
      \begin{pmatrix}
          \omega_o-\mu+jr_o & 0\\[4pt]
          0 & \omega_o+\mu+jr_o
      \end{pmatrix},
      \quad
      \bar{\bar{\mu}} = 
      \begin{pmatrix}
          0 & \delta\omega(t) \\[4pt]
          \delta\omega(t) & 0
      \end{pmatrix}.
    \end{aligned}\right.
\end{equation*}
One can notice that modulating term $\delta\omega(t)$ appears in off-diagonal elements of the coupling matrix $\bar{\bar{\mu}}$, indicating modulation-induced coupling between the supermodes.

The prior analysis considered the isolated system of coupled cavities which are not coupled to external waveguide(s). Now, we introduce coupling of the supermodes to input/output ports, which is characterized by energy amplitude decay rates $r_{e,s}$ and $r_{e,a}$ for the symmetric and the antisymmetric supermodes, respectively. The CMT equation for the amplitudes of the supermode energies and the input/output optical fields can be written as
\begin{equation}\label{EqnModeAmpExtCoupling}
	\left.\begin{aligned}
      &\frac{d}{dt} \bar{b}=j\bar{\bar{\omega}}\cdot\bar{b}-j\bar{\bar{\mu}}\cdot\bar{b}-j\bar{\bar{M_i}}\cdot\bar{s}_+ \\[6pt]
      &\bar{s}_- = -j\bar{\bar{M_o}}\cdot\bar{b} + \bar{s}_+
	\end{aligned}\right.
\end{equation}
where
\begin{equation*}
	\left.\begin{aligned}
      \quad
      &\bar{\bar{\omega}} = 
      \begin{pmatrix}
          \omega_o-\mu+j(r_o+r_{e,s}) & 0\\[4pt]
          0 & \omega_o+\mu+j(r_o+r_{e,a})
      \end{pmatrix},
      \quad
      \bar{\bar{\mu}} = 
      \begin{pmatrix}
          0 & \delta\omega(t)  \\[4pt]
          \delta\omega(t) & 0
      \end{pmatrix},  \\[6pt]
      &\bar{\bar{M_i}}=
      \begin{pmatrix}
          \sqrt{2r_{e,s}} & 0 \\[4pt]
          0 & \sqrt{2r_{e,a}}
      \end{pmatrix},
      \quad
      \bar{\bar{M_o}} = \bar{\bar{M_i}}^T,
      \quad
      \bar{s}_+ =
      \begin{pmatrix}
          s_{+1} \\[4pt]
          s_{+2}
      \end{pmatrix},
      \quad
      \bar{s}_- =
      \begin{pmatrix}
          s_{-1} \\[4pt]
          s_{-2}
      \end{pmatrix}.
    \end{aligned}\right.
\end{equation*}
Here, $s_{+1},\ s_{+2},\ s_{-1},\ s_{-2}$ are the input/output field amplitudes, shown next to corresponding ports in Figs.~\ref{fig:ModTypes}(c,d). In the proposed modulator the only optical input is the laser pump $s_{+1}$, therefore $s_{+2}$ is always assumed to be zero. The fields at the output of the modulator have amplitude $s_{-1}$ for the residual pump and amplitude $s_{-2}$ for the generated optical sideband.

In this work, we consider the basic and the generalized designs of Figs.~\ref{fig:ModTypes}(c,d) for physically realizing the described coupling of the supermodes to the input and the output ports. In the \textit{basic} design [Fig.~\ref{fig:ModTypes}(c)], the input pump light couples to the symmetric mode and the light in the optical sideband is extracted from the antisymmetric supermode via the same bus waveguide, which is coupled to one of the cavities. As mentioned in Secs.~\ref{sec:OpPrinc} and \ref{sec:OptResAnalysis}, the coupling rates of the supermodes, $r_{e,s}$ and $r_{e,a}$, are the same and are equal to half of the energy amplitude coupling rate of the cavity to the bus waveguide, i.e. $r_{e,s}=r_{e,a}=r_e/2$. In the \textit{generalized} design [Fig.~\ref{fig:ModTypes}(d)], the pump is coupled into the symmetric supermode and and the generated sideband is coupled out of the antisymmetric supermode through separate input and output waveguides. Each waveguide is coupled to both cavities with coupling strength described by cavity decay rates $r_{e,in}$ and $r_{e,out}$. As explained in Secs.~\ref{sec:OpPrinc} and \ref{sec:OptResAnalysis}, the supermode coupling rates $r_{e,s}$ and $r_{e,a}$ are set independently by the ring-waveguide coupling strengths $r_{e,in}$ and $r_{e,out}$, with $r_{e,s} = 2r_{e,in}$ and $r_{e,a} = 2r_{e,out}$, where the factors of $2$ result from constructive interference of the fields coupled out from the two rings. Note that the CMT loss rate $r_o$, the ring-to-waveguide coupling rates $r_{e,in}, r_{e,out}$, and the ring-to-ring coupling rate $\mu$ are related to the propagation loss $\alpha$ [dB/m], the waveguide-to-ring power coupling coefficients $\kappa_{in}^2, \kappa_{out}^2$, and the ring-to-ring power coupling coefficient $\kappa_{rr}^2$ through
\begin{equation}\label{EqnCavityParamConversion}
	\left.\begin{aligned}
      &\alpha=\frac{20n_gr_o}{ln(10)c},\quad\kappa_{in}^2=\frac{2r_{e,in}}{\Delta f_{FSR}},\quad\kappa_{out}^2=\frac{2r_{e,out}}{\Delta f_{FSR}},\quad\kappa_{rr}^2=\frac{\mu^2}{\Delta f_{FSR}^2},
	\end{aligned}\right.
\end{equation}
where $\Delta f_{FSR}$ is the free spectral range of the ring (in Hz) \cite{Little1997}. 

Equations (\ref{EqnModeAmp}) can be solved when an input harmonic modulating signal, $\delta\omega(t) = \frac{\delta\omega_m}{2}\cos(\Omega t)$, is applied, where $\delta\omega_m$ is peak-to-peak resonance frequency swing. Assuming that the frequency of the input pump wave is close to the frequency of the symmetric supermode $\omega_o-\mu$ and the frequency of the generated sideband is close to the frequency of the antisymmetric supermode, we can simplify the analysis by introducing slowly-varying envelopes with
\begin{equation*}
b_1(t)\equiv B_1(t)e^{j(\omega_o-\mu)t},\quad b_2(t)\equiv B_2(t)e^{j(\omega_o+\mu)t},\quad  s_{+1}(t)\equiv S_{+1}(t)e^{j(\omega_o-\mu)t}
\end{equation*}
where $B_1(t)$, $B_2(t)$ are the slowly varying envelopes of the supermode energy amplitudes, and $S_{+1}(t)$ is the slowly varying envelope of the input pump wave. The coupled mode equations (\ref{EqnModeAmp}) can then be rewritten as
\begin{equation}\label{EqnModeAmpEnvlp}
	\left.\begin{aligned}
      &\frac{d}{dt} \bar{B}=j\bar{\bar{H}}\cdot\bar{B}-j\bar{\bar{M_i}}\cdot\bar{S}_+ \\[6pt]
      &\bar{S}_- = -j\bar{\bar{M_o}}\cdot\bar{B} + \bar{S}_+
	\end{aligned}\right.
\end{equation}
where 
\begin{equation*}
	\left.\begin{aligned}
      &\bar{B} = 
      \begin{pmatrix}
          B_1 \\[4pt]
          B_2
      \end{pmatrix},
      \quad
      \bar{\bar{H}} = 
      \begin{pmatrix}
          j(r_o+r_{e,s}) & -\dfrac{\delta\omega_m}{4}(e^{j(\Omega+2\mu)t}+e^{-j(\Omega-2\mu)t})\\[6pt]
          -\dfrac{\delta\omega_m}{4}(e^{j(\Omega-2\mu)t}+e^{-j(\Omega+2\mu)t}) & j(r_o+r_{e,a})
      \end{pmatrix},\\[6pt]
      &\bar{S}_+ =
      \begin{pmatrix}
          S_{+1} \\[4pt]
          S_{+2}
      \end{pmatrix},
      \quad
      \bar{S}_- =
      \begin{pmatrix}
          S_{-1} \\[4pt]
          S_{-2}
      \end{pmatrix},
    \end{aligned}\right.
\end{equation*}

To allow closed form solutions, the contribution from the rapidly oscillating exponentials with arguments $\pm(\Omega+2\mu)$ can be neglected compared to the contribution from slowly varying exponentials with arguments $\pm(\Omega-2\mu)$. For numerical simulations of the CMT equations' evolution, they may be retained.

Using the convention introduced in Sec.~\ref{sec:OptResAnalysis} and Fig.~\ref{fig:ConvEffParams}, we let the pump frequency be different from the symmetric supermode frequency $\omega_o-\mu$ by $\Delta\omega_1$, so that $S_{+1}(t)=\tilde{S}_{+1}e^{j\Delta\omega_1t}$. We look for solutions for the slowly varying envelopes in the form of $B_1(t) = \tilde{B}_1e^{j\Delta\omega_1t}$ and $B_2(t) = \tilde{B}_2e^{j\Delta\omega_2t}$, where $\tilde{B}_1$ and $\tilde{B}_2$ are constants, and $\Delta\omega_2$ is the detuning of generated optical sideband frequency from the resonance frequency of antisymmetric supermode, as shown in Fig.~\ref{fig:ConvEffParams}. The relation between $\Delta\omega_1$, $\Delta\omega_2$, the frequency $\Omega$ and coupling rate $\mu$ is given by Eq.~(\ref{EqnDw2MinusDw1}). 

Substituting the expressions for $S_{+1}(t)$, $B_1(t)$, and $B_2(t)$ into Eq.~(\ref{EqnModeAmpEnvlp}) and replacing the derivatives with $j\Delta\omega_1$ and $j\Delta\omega_2$ we get
\begin{equation}\label{EqnModeAmpEnvlpTilde}
	\left.\begin{aligned}
      &j\bar{\bar{\Delta\omega}}\cdot\bar{\tilde{B}}=j\bar{\bar{H}}\cdot\bar{\tilde{B}}-j\bar{\bar{M_i}}\cdot\bar{\tilde{S}}_+ \\[6pt]
      &\bar{\tilde{S}}_- = -j\bar{\bar{M_o}}\cdot\bar{\tilde{B}} + \bar{\tilde{S}}_+
	\end{aligned}\right.
\end{equation}
where
\begin{equation*}
	\left.\begin{aligned}
      &\bar{\bar{\Delta\omega}} = 
      \begin{pmatrix}
          \Delta\omega_1 & 0\\[4pt]
          0 & \Delta\omega_2
      \end{pmatrix},
      \bar{\tilde{B}} = 
      \begin{pmatrix}
          \tilde{B}_1 \\[4pt]
          \tilde{B}_2
      \end{pmatrix},
      \bar{\bar{H}} = 
      \begin{pmatrix}
          j(r_o+r_{e,s}) & -\dfrac{\delta\omega_m}{4}\\[6pt]
          -\dfrac{\delta\omega_m}{4} & j(r_o+r_{e,a})
      \end{pmatrix},
      \bar{\tilde{S}}_+ =
      \begin{pmatrix}
          \tilde{S}_{+1} \\[4pt]
          \tilde{S}_{+2}
      \end{pmatrix},
      \bar{\tilde{S}}_-
      \begin{pmatrix}
          \tilde{S}_{-1} \\[4pt]
          \tilde{S}_{-2}
      \end{pmatrix}.
    \end{aligned}\right.
\end{equation*}
The first equation in (\ref{EqnModeAmpEnvlpTilde}) can be readily solved for the supermode envelopes $\tilde{B}_1$ and $\tilde{B}_2$, and the amplitude of the optical sideband $\tilde{S}_{-2}$ can be found from the second equation in (\ref{EqnModeAmpEnvlpTilde}). The conversion efficiency of the modulator, defined as the ratio of the output power at optical sideband to the input laser power according to Eq.~(\ref{EqnConvEffDef}), is then
\begin{flalign}\label{EqnConvEffGeneral}
	&\quad\ G = \left\lvert\frac{\tilde{S}_{-2}}{\tilde{S}_{+1}}\right\rvert^2=&
\end{flalign}
\begin{equation*}
 	\left.\begin{aligned}
     &=\dfrac{\frac{1}{4}r_{e,s} r_{e,a}\delta\omega_m^2}{\left[(r_o+r_{e,a})\Delta\omega_1+(r_o+r_{e,s})\Delta\omega_2\right]^2+\left[(r_o+r_{e,s})(r_o+r_{e,a})+\left(\dfrac{\delta\omega_m}{4}\right)^2-\Delta\omega_1\Delta\omega_2\right]^2}
 	\end{aligned}\right.
\end{equation*}
The expression (\ref{EqnConvEffGeneral}) is used in this work for analyzing the operation of the proposed modulator. 

It is important to note that for weak modulation, i.e. $\delta \omega_m \ll 2r_o$, Eq.~(\ref{EqnConvEffGeneral}) can be written as
\begin{equation}\label{EqnConvEffSplitResContribution}
	\left.\begin{aligned}
      &G =\left(\frac{\delta\omega_m}{4}\right)^2\times\frac{2r_{e,s}}{\Delta\omega_1^2 +(r_o+r_{e,s})^2}\times\frac{2r_{e,a}}{\Delta\omega_2^2 +(r_o+r_{e,a})^2},
	\end{aligned}\right.
\end{equation}
where the second and the third fractions can be recognized as the Lorentzian lineshapes of the symmetric and the antisymmetric supermodes evaluated at the frequencies of the optical carrier and the sideband, respectively. This confirms the idea that for efficient sideband conversion, each of the interacting optical waves need to be resonant in the device.

\section*{Appendix B: List of symbols}
\setcounter{table}{0}
\renewcommand{\thetable}{B.\arabic{table}}
\vspace{-0.55cm}
\begin{center}
    \arrayrulecolor{black}
\begin{longtable}{l l}
    \caption{List of the commonly used symbols and their description.}
    \label{table:ParamDescript}\\
	\hline
    \textbf{Symbol} & \textbf{Description} \\ \hline\hline
    \endfirsthead 
    \hline
    \endfoot
    \hline
    \textbf{Symbol} & \textbf{Description} \\ \hline\hline
    \endhead
    \hline
    \endlastfoot
    $\omega_o$ & Resonance frequency of the optical cavities when uncoupled. \\[0pt]
    $r_o$ & Intrinsic decay rate of the cavity and supermode energy amplitudes due to losses. \\[0pt]
    $\mu$ & Coupling rate between the optical cavities.\\[0pt]
    $r_{e}$ & Coupling rate from the bottom cavity to the bus waveguide in the basic design.\\[0pt]
    $r_{e,in}$ & Coupling rate from the cavities to the input waveguide in the generalized design.\\[0pt]
    $r_{e,out}$ & Coupling rate from the cavities to the output waveguide in the generalized design.\\[0pt]
    $r_{e,s}$ & Coupling rate from the symmetric supermode to the input/output waveguide(s).\\[0pt]
    $r_{e,a}$ & Coupling rate from the antisymmetric supermode to the input/output waveguide(s).\\[0pt]
    $\delta\omega_m$ & Peak-to-peak cavity resonance frequency swing due to modulation.\\[0pt]
    $\Delta\omega_1$ & Detuning of the laser pump from the symmetric resonance in frequency.\\[0pt]
    $\Delta\omega_2$ & Detuning of the optical sideband from the antisymmetric resonance in frequency.\\[0pt]
    $\Omega_o$ & Carrier frequency of the RF drive signal and frequency of the RF resonance.\\[0pt]
    $\Omega$ & Arbitrary frequency of the RF drive signal.\\[0pt]
    $\Delta\Omega_{3dB}$ & Photon-lifetime-limited RF bandwidth.\\[0pt]
    $\alpha$ & Waveguide propagation loss of the optical cavities. \\[0pt]
    $V_{\pi}L$ & Voltage-length product of the electro-optic phase shifters of the cavities. \\[0pt]
    $Z_o$ & Characteristic impedance of the transmission line delivering the RF signal.\\[0pt]
    $C_m$ & Capacitance of the electro-optic region of the active cavities.\\[0pt]
    $R_m$ & Parasitic resistance of the active cavities.\\[0pt]
    $L_1$ & Inductance of the monolithically integrated inductors.\\[0pt]
    $R_{L_1}$ & Parasitic resistance of the inductor $L_1$.\\[0pt]
 	$Q_L$ & Quality factor of the inductors. \\[0pt]
    $Q_{RF}^{tot}$ & Total quality factor of the RF resonator.\\
\end{longtable}
    \vspace{-1.0cm}
\end{center}

\section*{Acknowledgements}
This work was supported in part by Ball Aerospace and Technologies Corp. and by NSF Award \#1701596. We thank Todd Pett of Ball Aerospace for his input and discussions.


\bibliography{DRrefs}

\begin{thebibliography}{10}

\bibitem{Capmany2007}
J.~Capmany and D.~Novak, ``{Microwave photonics combines two worlds},'' {\em
  Nature Photonics}, vol.~1, no.~6, pp.~319--330, 2007.

\bibitem{Marpaung2013}
D.~Marpaung, C.~Roeloffzen, R.~Heideman, A.~Leinse, S.~Sales, and J.~Capmany,
  ``{Integrated microwave photonics},'' {\em Laser {\&} Photonics Reviews},
  vol.~7, pp.~506--538, jul 2013.

\bibitem{Minasian2006}
R.~A. Minasian, ``{Photonic signal processing of microwave signals},'' {\em
  IEEE Transactions on Microwave Theory and Techniques}, vol.~54, no.~2,
  pp.~832--846, 2006.

\bibitem{Pett2018}
T.~Pett, J.~Lee, Y.~Ehrlichman, H.~Gevorgyan, A.~Khilo, and M.~A.
  Popovi{\'{c}}, ``{Photonics-based Microwave Radiometer for Hyperspectral
  Earth Remote Sensing},'' in {\em Proceedings of IEEE International Topical
  Meeting on Microwave Photonics}, IEEE, 2018.

\bibitem{Williamson2001}
R.~C. Williamson, ``{Sensitivity-bandwidth product for electro-optic
  modulators},'' {\em Optics Letters}, vol.~26, p.~1362, sep 2001.

\bibitem{poon2008}
W.~D. Sacher and J.~K.~S. Poon, ``{Dynamics of microring resonator
  modulators.},'' {\em Optics Express}, vol.~16, no.~20, pp.~15741--15753,
  2008.

\bibitem{Ehrlichman2018}
Y.~Ehrlichman, A.~Khilo, and M.~A. Popovi{\'{c}}, ``{Optimal design of a
  microring cavity optical modulator for efficient RF-to-optical conversion},''
  {\em Optics Express}, vol.~26, p.~2462, feb 2018.

\bibitem{levi2001}
D.~A. Cohen, M.~Hossein-Zadeh, and A.~F. Levi, ``{High-Q microphotonic
  electro-optic modulator},'' {\em Solid-State Electronics}, vol.~45, no.~9,
  pp.~1577--1589, 2001.

\bibitem{maleki2003}
V.~S. Ilchenko, A.~A. Savchenkov, A.~B. Matsko, and L.~Maleki,
  ``{Whispering-gallery-mode electro-optic modulator and photonic microwave
  receiver},'' {\em Journal of the Optical Society of America B}, vol.~20,
  no.~2, p.~333, 2003.

\bibitem{lipson2014}
L.~D. Tzuang, M.~Soltani, Y.~H.~D. Lee, and M.~Lipson, ``{High RF carrier
  frequency modulation in silicon resonators by coupling adjacent
  free-spectral-range modes},'' {\em Optics Letters}, vol.~39, no.~7,
  pp.~1799--1802, 2014.

\bibitem{wade2014}
M.~T. Wade, X.~Zeng, and M.~A. Popovi{\'{c}}, ``{Wavelength conversion in
  modulated coupled-resonator systems and their design via an equivalent linear
  filter representation.},'' {\em Optics Letters}, vol.~40, no.~1, pp.~107--10,
  2015.

\bibitem{Ehrlichman2016}
Y.~Ehrlichman, N.~Dostart, and M.~A. Popovi{\'{c}}, ``{Dual-cavity resonant
  modulators for efficient narrowband RF/microwave photonics},'' in {\em
  Proceedings of International Topical Meeting on Microwave Photonics},
  pp.~165--168, IEEE, 2016.

\bibitem{Ehrlichman2017}
Y.~Ehrlichman and M.~A. Popovi{\'{c}}, ``{Dual-Cavity Optically and
  Electrically Resonant Modulators for Efficient Narrowband RF/Microwave
  Photonics},'' in {\em Conference on Lasers and Electro-Optics}, no.~paper
  JW2A.115, Optical Society of America, 2017.

\bibitem{loncar2017}
M.~Soltani, M.~Zhang, C.~Ryan, G.~J. Ribeill, C.~Wang, and M.~Loncar,
  ``{Efficient quantum microwave-to-optical conversion using electro-optic
  nanophotonic coupled resonators},'' {\em Physical Review A}, vol.~96,
  p.~043808, oct 2017.

\bibitem{Gevorgyan2018}
H.~Gevorgyan, A.~Khilo, and M.~A. Popovi{\'{c}}, ``{Broadband efficient
  coupled-cavity electro-optic modulators based on Q engineering for RF
  photonics applications},'' in {\em Conference on Lasers and Electro-Optics},
  no.~paper JW2A.59, Optical Society of America, 2018.

\bibitem{Lee2004}
T.~H. Lee, {\em {The design of CMOS radio-frequency integrated circuits}}.
\newblock Cambridge University Press, 2004.

\bibitem{Stojanovic2018}
V.~Stojanovi{\'{c}}, R.~J. Ram, M.~Popovi{\'{c}}, S.~Lin, S.~Moazeni, M.~Wade,
  C.~Sun, L.~Alloatti, A.~Atabaki, F.~Pavanello, N.~Mehta, and P.~Bhargava,
  ``{Monolithic silicon-photonic platforms in state-of-the-art CMOS SOI
  processes [Invited]},'' {\em Optics Express}, vol.~26, p.~13106, may 2018.

\bibitem{45rfsoi2017}
GlobalFoundries, ``{Advanced 45nm RF SOI technology},''
  No.~https://www.globalfoundries.com/sites/default/files/pb-45rfsoi-en.pdf.

\bibitem{Lee2007}
S.~Lee, B.~Jagannathan, S.~Narasimha, A.~Chou, N.~Zamdmer, J.~Johnson,
  R.~Williams, L.~Wagner, J.~Kim, J.-O. Plouchart, J.~Pekarik, S.~Springer, and
  G.~Freeman, ``{Record RF performance of 45-nm SOI CMOS Technology},'' in {\em
  International Electron Devices Meeting}, pp.~255--258, IEEE, dec 2007.

\bibitem{plant2015}
D.~Patel, S.~Ghosh, M.~Chagnon, A.~Samani, V.~Veerasubramanian, M.~Osman, and
  D.~V. Plant, ``{Design, analysis, and transmission system performance of a 41
  GHz silicon photonic modulator},'' {\em Optics Express}, vol.~23, p.~14263,
  jun 2015.

\bibitem{hochberg2014}
R.~Ding, Y.~Liu, Y.~Ma, Y.~Yang, Q.~Li, A.~E.~J. Lim, G.~Q. Lo, K.~Bergman,
  T.~Baehr-Jones, and M.~Hochberg, ``{High-speed silicon modulator with
  slow-wave electrodes and fully independent differential drive},'' {\em
  Journal of Lightwave Technology}, vol.~32, no.~12, pp.~2240--2247, 2014.

\bibitem{miyazawa1998}
K.~Noguchi, O.~Mitomi, and H.~Miyazawa, ``{Millimeter-wave Ti:LiNbO3 optical
  modulators},'' {\em Journal of Lightwave Technology}, vol.~16, no.~4,
  pp.~615--619, 1998.

\bibitem{mcgee2002}
M.~Lee, H.~E. Katz, C.~Erben, D.~M. Gill, P.~Gopalan, J.~D. Heber, and D.~J.
  McGee, ``{Broadband modulation of light by using an electro-optic polymer},''
  {\em Science}, vol.~298, no.~5597, pp.~1401--1403, 2002.

\bibitem{leuthold2017}
C.~Hoessbacher, A.~Josten, B.~Baeuerle, Y.~Fedoryshyn, H.~Hettrich, Y.~Salamin,
  W.~Heni, C.~Haffner, C.~Kaiser, R.~Schmid, D.~L. Elder, D.~Hillerkuss,
  M.~M{\"{o}}ller, L.~R. Dalton, and J.~Leuthold, ``{Plasmonic modulator with
  {\textgreater}170 GHz bandwidth demonstrated at 100 GBd NRZ},'' {\em Optics
  Express}, vol.~25, p.~1762, feb 2017.

\bibitem{Xu2005}
Q.~Xu, B.~Schmidt, S.~Pradhan, and M.~Lipson, ``{Micrometre-scale silicon
  electro-optic modulator},'' {\em Nature}, vol.~435, pp.~325--327, may 2005.

\bibitem{moazeni2017}
S.~Moazeni, S.~Lin, M.~Wade, L.~Alloatti, R.~J. Ram, M.~A. Popovi{\'{c}}, and
  V.~Stojanovi{\'{c}}, ``{A 40-Gb/s PAM-4 Transmitter Based on a Ring-Resonator
  Optical DAC in 45-nm SOI CMOS},'' {\em IEEE Journal of Solid-State Circuits},
  vol.~52, no.~12, pp.~3503--3516, 2017.

\bibitem{watts2014}
E.~Timurdogan, C.~M. Sorace-Agaskar, J.~Sun, E.~{Shah Hosseini}, A.~Biberman,
  and M.~R. Watts, ``{An ultralow power athermal silicon modulator},'' {\em
  Nature Communications}, vol.~5, p.~4008, dec 2014.

\bibitem{sun2015}
C.~Sun, M.~T. Wade, Y.~Lee, J.~S. Orcutt, L.~Alloatti, M.~S. Georgas, A.~S.
  Waterman, J.~M. Shainline, R.~R. Avizienis, S.~Lin, B.~R. Moss, R.~Kumar,
  F.~Pavanello, A.~H. Atabaki, H.~M. Cook, A.~J. Ou, J.~C. Leu, Y.~H. Chen,
  K.~Asanovi{\'{c}}, R.~J. Ram, M.~A. Popovi{\'{c}}, and V.~M.
  Stojanovi{\'{c}}, ``{Single-chip microprocessor that communicates directly
  using light},'' {\em Nature}, vol.~528, no.~7583, pp.~534--538, 2015.

\bibitem{campenhout2015}
M.~Pantouvaki, P.~Verheyen, J.~{De Coster}, G.~Lepage, P.~Absil, and J.~{Van
  Campenhout}, ``{56Gb/s ring modulator on a 300mm silicon photonics
  platform},'' in {\em European Conference on Optical Communication}, pp.~1--3,
  IEEE, sep 2015.

\bibitem{Yu2014}
H.~Yu, M.~Pantouvaki, P.~Verheyen, G.~Lepage, P.~Absil, W.~Bogaerts, and
  J.~{Van Campenhout}, ``{Silicon dual-ring modulator driven by differential
  signal},'' {\em Optics Letters}, vol.~39, no.~22, pp.~6379--6382, 2014.

\bibitem{Savchenkov2010}
A.~A. Savchenkov, A.~B. Matsko, W.~Liang, V.~S. Ilchenko, D.~Seidel, and
  L.~Maleki, ``{Single-Sideband Electro-Optical Modulator and Tunable Microwave
  Photonic Receiver},'' {\em IEEE Transactions on Microwave Theory and
  Techniques}, vol.~58, pp.~3167--3174, nov 2010.

\bibitem{Little1997}
B.~Little, S.~Chu, H.~Haus, J.~Foresi, and J.-P. Laine, ``{Microring resonator
  channel dropping filters},'' {\em Journal of Lightwave Technology}, vol.~15,
  pp.~998--1005, jun 1997.

\bibitem{Zeng2015}
X.~Zeng, C.~M. Gentry, and M.~A. Popovi{\'{c}}, ``{Four-wave mixing in silicon
  coupled-cavity resonators with port-selective, orthogonal supermode
  excitation},'' {\em Optics Letters}, vol.~40, p.~2120, may 2015.

\bibitem{Wang2018}
C.~Wang, M.~Zhang, B.~Stern, M.~Lipson, and M.~Lon{\v{c}}ar, ``{Nanophotonic
  lithium niobate electro-optic modulators},'' {\em Optics Express}, vol.~26,
  p.~1547, jan 2018.

\bibitem{Jiang2000}
{Hongrui Jiang}, J.-L. Yeh, {Ye Wang}, and {Norman Tien},
  ``{Electromagnetically shielded high-Q CMOS-compatible copper inductors},''
  in {\em International Solid-State Circuits Conference}, pp.~330--331, IEEE,
  2000.

\bibitem{Dickson2005}
T.~Dickson, M.-A. LaCroix, S.~Boret, D.~Gloria, R.~Beerkens, and S.~Voinigescu,
  ``{30-100-GHz inductors and transformers for millimeter-wave (Bi)CMOS
  integrated circuits},'' {\em IEEE Transactions on Microwave Theory and
  Techniques}, vol.~53, pp.~123--133, jan 2005.

\bibitem{Alexander2017}
K.~Alexander, J.~P. George, B.~Kuyken, J.~Beeckman, and D.~{Van Thourhout},
  ``{Broadband Electro-optic Modulation using Low-loss PZT-on-Silicon Nitride
  Integrated Waveguides},'' in {\em Conference on Lasers and Electro-Optics},
  no.~paper JTh5C.7, Optical Society of America, 2017.

\bibitem{Eltes2017}
F.~Eltes, M.~Kroh, D.~Caimi, C.~Mai, Y.~Popoff, G.~Winzer, D.~Petousi,
  S.~Lischke, J.~E. Ortmann, L.~Czornomaz, L.~Zimmermann, J.~Fompeyrine, and
  S.~Abel, ``{A novel 25 Gbps electro-optic Pockels modulator integrated on an
  advanced Si photonic platform},'' in {\em International Electron Devices
  Meeting}, pp.~24.5.1--24.5.4, IEEE, dec 2017.

\bibitem{Kieninger2018}
C.~Kieninger, Y.~Kutuvantavida, D.~L. Elder, S.~Wolf, H.~Zwickel, M.~Blaicher,
  J.~N. Kemal, M.~Lauermann, S.~Randel, W.~Freude, L.~R. Dalton, and C.~Koos,
  ``{Ultra-high electro-optic activity demonstrated in a silicon-organic hybrid
  modulator},'' {\em Optica}, vol.~5, p.~739, jun 2018.

\bibitem{Yong2017}
Z.~Yong, W.~D. Sacher, Y.~Huang, J.~C. Mikkelsen, Y.~Yang, X.~Luo, P.~Dumais,
  D.~Goodwill, H.~Bahrami, P.~G.-Q. Lo, E.~Bernier, and J.~K.~S. Poon,
  ``{Efficient Single-Drive Push-Pull Silicon Mach-Zehnder Modulators with
  U-Shaped PN Junctions for the O-Band},'' in {\em Optical Fiber Communication
  Conference}, no.~paper Tu2H.2, Optical Society of America, 2017.

\end{thebibliography}
\bibliographystyle{ieeetr}
\end{document}